\documentclass[12pt]{article}

\usepackage{ragged2e} 
\usepackage{changepage}

\usepackage{siunitx} % degree symbol
\usepackage{amsmath}
\usepackage{amssymb}
\usepackage{xfrac} % for other fraction (e.g. sfrac)
\usepackage{bbm}

\usepackage{graphicx}
\usepackage{subcaption}
\usepackage{sidecap}
\usepackage{wrapfig}

\usepackage{caption}
\usepackage{verbatim}
\usepackage[english]{babel}

\usepackage{array}
\usepackage{adjustbox}
\usepackage{float}
\usepackage{enumerate}
\usepackage{multirow}
\usepackage{multicol}
\usepackage[table,xcdraw]{xcolor}
\usepackage[round]{natbib}

\usepackage{csvsimple}
\usepackage{bbm}
\usepackage{pdflscape}
\usepackage{breqn}
\usepackage{lscape}

\usepackage{amsfonts}
\usepackage{latexsym}
 \usepackage[titletoc]{appendix}  
\usepackage[colorlinks=true, citecolor= blue]{hyperref} % hyperlinks
\hypersetup{linkcolor=black} % hyperlinks
\usepackage[all]{hypcap} % hyperlinks

\usepackage[a4paper,top=2cm,bottom=2cm,left=3cm,right=3cm,marginparwidth=1.75cm]{geometry}

\usepackage{float} 
\usepackage{xcolor}
\usepackage{setspace}
\usepackage{hyperref}

\usepackage{authblk}

\newcommand{\shortsmile}[1]{\overset{\scalebox{0.6}[0.6]{$\smile$}}{#1}\mspace{-1mu}}

\title{\vspace{-1cm} \huge \textbf{Spatially varying distributed lag non-linear models using Laplacian P-splines}}
\date{}

\author[1]{Sara Rutten
\footnote{Corresponding author. {\textit{E-mail address}: sara.rutten@uhasselt.be}}}
\author[1,2]{Thomas Neyens}
\author[1]{Elisa Duarte}
\author[3]{Antonio Gasparrini}
\author[1]{Christel Faes}

\affil[1]{Interuniversity Institute for Biostatistics and statistical Bioinformatics (I-BioStat), Data Science Institute (DSI), Hasselt University, Hasselt, Belgium}
\affil[2]{L-BioStat, Department of Public Health and Primary Care, KU Leuven, Leuven, Belgium}
\affil[3]{Environment \& Health Modelling (EHM) Lab, Department of Public Health Environments and Society, London School of Hygiene \& Tropical Medicine, London, United Kingdom}

\begin{document}
\maketitle

\begin{abstract}
Although distributed lag non-linear models (DLNMs) are commonly used to quantify delayed and non-linear exposure-response relationships, most existing applications assume that these relationships are constant across space. However, in many geographical and environmental studies, local characteristics vary substantially across areas, making a spatially varying effect more realistic. 

Extending DLNMs to allow for spatial heterogeneity remains challenging, and only a limited number of modelling strategies have been proposed in literature. The most popular extension is a two-stage meta-analysis approach, which requires sufficiently large sample sizes at each location. Therefore, its usefulness is limited when working with sparse count data in small area data analyses. Although a number of alternative one-stage approaches have been introduced, their computational burden restricts their applicability in real-life data applications. 

In this paper, we introduce a computationally efficient Bayesian one-stage spatially-varying DLNM for count data. We define four model variants, differing in the assumed spatial dependence structure and the flexibility of the DLNM spline specification. To address the computational burden typically associated with these flexible models, we use Laplace approximations, offering an efficient alternative to classically used Markov Chain Monte Carlo (MCMC) approaches. Model comparison criteria are provided to facilitate the selection of a suitable model in a real-life data application. 

The proposed methods are evaluated through simulation studies, and their practical usefulness is illustrated through a real-life data application, investigating the temperature-mortality relationship in every municipality of Sicily, Italy.
  
  
 \thispagestyle{empty}
\end{abstract}

\newpage
\clearpage\null

\pagenumbering{arabic}

\section{Introduction}
Environmental exposures such as air pollution, heat and drought can pose both short- and long-term effects on human health and ecological systems. For example, \cite{Rutten2025} identified both immediate and delayed effects (up to $8$ weeks) of black carbon pollution on COVID-19 incidences in Belgium. Moreover, \cite{Redana2024} found that temperature has a lagged effect over $3$ weeks on thermal tolerance in mayflies and \cite{Xu2024} identified lagged effects of several months when analysing the impact of drought on vegetation health. As the frequency of extreme weather events increases, understanding not only the direct but also the delayed effects of these exposures has become increasingly relevant. Many exposures show effects that last several days or weeks and failing to capture these lagged patterns can result in substantial underestimation of environmental risks \citep{gasparrini2010}. 

Standard methods for modelling delayed effects of exposures have traditionally relied on restrictive assumptions that limit how lagged relationships can be represented. For example, in moving-average models, all lagged information is collapsed into a single summary measure, which prevents the estimation of lag-specific exposure effects. \cite{gasparrini2010} advanced this field substantially by the generalization of distributed lag non-linear models (DLNMs), which provide a powerful framework to model exposure-lag-response relationships. These models are well-suited to capture complex relationships, since the use of spline basis functions allows for a smooth effect in both the lag and exposure dimensions. To further enhance the model's flexibility, \cite{Gasparrini2017} combined the DLNM with a penalized spline framework \citep{Eilers1996}, allowing the model to adapt to complex patterns while still avoiding overfitting. Because of its flexibility, the DLNM became a popular modelling method in environmental health research \citep{Gasparrini2015, Bao2016} but is also frequently applied in other scientific fields like infectious-disease modelling \citep{Fajgenblat2024, Rutten2025} and ecology \citep{Redana2024, Xu2024}.

Despite the popularity of the DLNM framework, its use in multi-location studies continues to present a challenge. In particular, exposure-lag-response relationships might vary spatially due to (unobserved) area-specific covariates, which is not well accommodated by the currently available methodology. For instance, \cite{choi2022} argued that the temperature-mortality relationship is modified by the amount of greenspace in a city. Moreover \cite{Zheng2023} found that the effect of several environmental variables on diabetes mortality differs between coastal and inland cities. In \cite{Rutten2026_confounding}, different methods for modelling effect modification within a DLNM framework were compared. However, while these methods are useful when relevant covariates are available, they rely on the assumption that all the differences across locations can be explained by effect modification by a known covariate. Consequently, there is a growing need for approaches that allow a more data-driven understanding of spatially varying effects to identify previously unrecognized patterns. This issue is increasingly important in environmental health, where current insights are often derived from spatially aggregated analyses or stratification in analyst-specified regions, potentially masking fine-scale heterogeneity. 

To account for (unexplained) inherent spatial variation in the exposure-lag-response relationship between locations, \cite{Gasparrini2013} proposed a two-stage approach. In the first stage, a separate DLNM is independently fitted for every location. Afterwards, a meta-analysis is used to pool the coefficients across locations. However, although dependence between geographically close locations is often a realistic assumption, a classical meta-analysis does not take into account spatial correlation. \cite{Masselot2025} proposed an extension that allows for spatial predictions at unobserved locations by extrapolating residuals from observed locations using a geostatistical method. However, because this extension is applied after the meta-analysis, it only aims to provide spatial prediction, not directly informing or stabilizing estimates at the observed locations themselves. Hence, it is not well suited for small-area count-data analyses, where the primary goal is often to improve the robustness of the estimates in pre-specified areas, rather than to perform spatial extrapolation. Another spatial two-stage approach was introduced by \cite{Wang2022}. They proposed a (multivariate) CAR second-stage model to spatially smooth the first-stage relative risk (RR) estimates. When a wide range of exposure values is of interest, a separate univariate CAR model for each value is recommended to avoid the computational cost of a multivariate structure. However, this ignores the correlation structure between the RRs at different exposure levels. Alternatively, a multivariate structure can be assumed on the DLNM regression coefficients directly but, MCMC sampling is required, resulting in high computation times. 

An alternative single-stage hierarchical method has been proposed by \cite{Economou2025}. In this framework, a global exposure-lag surface is constructed alongside an area-specific deviation. For this area-specific effect, different modelling methods are proposed, all derived from an interaction between the DLNM crossbasis and an area-specific effect. The first option is to represent a spatial effect by an isotropic thin-plate regression spline of the coordinates. Alternatively, interactions with a Markov Random Field \citep{RueHeld2005} can be defined or, in the absence of a spatial structure, an area-specific unstructured random effect can be used. This approach allows borrowing strength across locations, resulting in more stable small-area estimates. However, the method, fitted in the \texttt{mgcv} package in R, can be computationally demanding, especially with a high number of areas. Although \cite{Economou2025} pointed out how GAMs can be viewed within a Bayesian framework, the method is not inherently Bayesian and does not allow the specification of a prior on the hyperparameters in the model.

Another small-area single-stage method that allows borrowing information within the DLNM framework has been proposed by \cite{Quijal-Zamorano2024}, using intrinsic CAR models \citep{besag1974} on the model coefficients. This method is based on MCMC sampling, resulting in very high computation times. However, \cite{Quijal-Zamorano2025} proposed an alternative Integrated Nested Laplace Approximation (INLA) specification of the model, highly decreasing the computational burden. Nonetheless, the method does not allow the use of penalized splines, requiring the user to choose the number of knots and the placement of the knots. These choices can strongly influence the results. \cite{Chen2025} proposed another scalable INLA alternative to the model of \cite{Quijal-Zamorano2024}. However, their model cannot be considered a true extension of the DLNM, as it does not allow for the estimation of the delayed effects, considerably limiting the models flexibility.

Together, existing approaches either lack spatial borrowing of strength, suffer from prohibitive computational demands, or fail to accommodate penalized splines and delayed effects within a unified hierarchical framework. This leaves a clear methodological gap for a computationally efficient, fully Bayesian, single-stage DLNM that supports location-specific exposure–lag–response estimation. To address this gap, we propose a novel single-stage Bayesian extension to the DLNM that allows for a location-specific (penalized) exposure-lag-response relationship in small- and large-area count studies. Our contribution is novel at it combines (i) a fully Bayesian hierarchical specification, (ii) penalized spline–based DLNMs, (iii) spatial borrowing of strength, and (iv) computational efficiency through Laplace approximation within a single unified framework.  The remainder of the paper is structured as follows. In Section~2, we introduce the novel methodological framework underlying this method. In Section~3, we evaluate its performance in comparison with the two-stage approach of \cite{Gasparrini2013}, considering both estimation accuracy and computational efficiency. Finally, Section~4 illustrates the applicability of the method in a real-world data application.

\section{Methodology}
\subsection{Common effect distributed lag non-linear model}
\label{sec:common}
Consider the response $y_{t,j}$ at time $t=1,\ldots,T$ and location $j=1,\ldots, J$. Under the assumption of a common exposure-lag-response relationship and assuming that counts $y_{t,j}$ follow a Poisson or negative binomial distribution, the mean $\mu_{t,j}=\mathbb{E}(y_{t,j})$ can be modelled through:
\begin{align}
\label{eq:DLNM_LPS}
    \log(\mu_{t,j}) = \beta_0+\sum_{h=1}^{H}{\beta_h a_{t,j,h}}+ s(x_{t,j}\ldots x_{t-L,j};\boldsymbol{\theta})+u_j,
\end{align}
with intercept $\beta_0$, $a_{t,j,h}$ representing linear covariate $a_h$ at time $t$ and location $j$ and $\beta_h$ the associated coefficient \citep{Rutten2026}. The smooth function $s(x_{t,j}\ldots x_{t-L,j};\boldsymbol{\theta})$, labelled as \textit{cross-basis}, describes the delayed effect of exposure $x$, using parameters $\boldsymbol{\theta} = (\theta_{11}, \ldots, \theta_{1v_l},\ldots \theta_{v_xv_l})^\top$, and $u_j$ is assumed to be a location-specific random effect. Following \cite{gasparrini2010}, the cross-basis function $s(x_{t,j}\ldots x_{t-L,j};\boldsymbol{\theta})$ can be modelled as:
\begin{align}
\label{eqn:DLNM}
    s(x_{t,j}, \ldots , x_{t-L,j} ; \boldsymbol{\theta})&=\sum_{i=1}^{v_x} \sum_{k=1}^{v_l}\left(\sum_{l=0}^L\widetilde{b}_i(x_{t-l,j})\shortsmile{b}_k(l) \right)\theta_{ik}=\boldsymbol{w_{t,j}^\top\theta},
\end{align}
with $\{\widetilde{b}_i\}_{i=1}^{v_x}$ and $\{\shortsmile{b}_k\}_{k=1}^{v_l}$ a spline basis for the exposure $(x_{t-l,j})$ and lag $(l)$ dimensions, respectively. The cross-basis vector $\boldsymbol{w_{t,j}}$ is obtained by applying the $v_x \cdot v_l$ cross-basis functions to $x_t$. More specifically, denote by $\tilde{B}_{t,j}$ the $(L+1) \times v_x$ dimensional basis matrix constructed by applying $\{\widetilde{b}_i\}_{i=1}^{v_x}$ to vector $\boldsymbol{q_{t,j}} = (x_{t,j}, x_{t-1,j}, \ldots, x_{t-L,j})^\top$. Moreover, denote by $\shortsmile{B}$ the $(L+1) \times v_l$ dimensional basis matrix constructed by applying $\{\shortsmile{b}_k\}_{k=1}^{v_l}$ to $\boldsymbol{l} = (0,1,\ldots,L)^\top$. Then, $\boldsymbol{w_{t,j}}^\top$ can be defined as $\left( \mathbf{1}^{\top}_{L + 1} \, A_{t,j} \right)$ with $\mathbf{1}$ a vector of ones and $A_{t,j} = \left( \widetilde{B}_{t,j} \otimes 1_{v_l}^\top \right) \odot \left( 1_{v_x}^\top \otimes \shortsmile{B} \right)$, where $\otimes$ and $\odot$ represent the Kronecker and Hadamard products, respectively. Natural splines are commonly used basis functions, as they impose boundary constraints, yielding stable behaviour near the edges \citep{Perperoglou2019}. However, the penalized version of the DLNM introduced by \cite{Gasparrini2017} offers a more flexible yet robust alternative. In this approach, P-splines \citep{Eilers1996} are used to flexibly capture complex relationships, with a penalty term included to avoid overfitting, even when many knots are specified. 

In a Bayesian framework, these penalties translate into priors on the coefficients $\boldsymbol{\theta}$ of a B-spline basis \citep{lang2004}. Following \cite{Rutten2026}, the penalty matrix can be constructed as $\mathcal{P}(\boldsymbol{\lambda})=\lambda_x (S_x\otimes I_{v_l}) + \lambda_l (I_{v_x} \otimes S_l)$, with $\boldsymbol{\lambda}=(\lambda_x,\lambda_l)^\top$ denoting the smoothing parameters for the exposure and lag dimension, respectively. The component penalty matrices are given by $S_x=D_{v_x}^{\top}D_{v_x} + \delta I_{v_x}$ and $S_{l}=D_{v_l}^{\top}D_{v_l} + \delta I_{v_l}$, where $D_{v_x}$ and $D_{v_l}$ are difference matrices of order $m$, commonly assumed $m=2$, and $\delta$ is a small ridge parameter ensuring numerical stability. The Kronecker structure ensures that smoothing is applied independently across exposure and lag directions. Thus, while the prior distribution of $\boldsymbol{\theta}$ in a natural spline basis is simply $\boldsymbol{\theta} \sim \mathcal{N}(0,\zeta^{-1} I)$ with $\zeta$ a small number (e.g. $\zeta = 10^{-5}$), in a penalized framework using Bayesian P-splines, this becomes the structured prior $$\boldsymbol{\theta} \sim \mathcal{N}(0,\mathcal{P}(\boldsymbol{\lambda})^{-1}).$$ Following \cite{Gasparrini2017} and \cite{Rutten2026}, an additional penalty in the lag dimension is recommended, encouraging shrinkage of the effect at longer lags. This penalty is particularly helpful in determining the correct lag period when the pre-specified maximum lag $L$ substantially exceeds the true lag period \citep{Obermeier2015}.

\subsection{Spatially varying distributed lag non-linear model}
\label{sec:Model_formulation}
In a multi-location study, the assumption of a single shared exposure–lag–response function $s(x_{t,j}\ldots x_{t-L,j};\boldsymbol{\theta})$ may be too restrictive. To accommodate spatial heterogeneity, we extend the model by allowing each location to deviate from the global DLNM surface. For location $j$, this can be written as:
\begin{align*}
    \log(\mu_{t,j}) = \beta_0+\sum_{h=1}^{H}{\beta_h a_{t,j,h}}+ s(x_{t,j}\ldots x_{t-L,j};\boldsymbol{\theta})+s(x_{t,j}\ldots x_{t-L,j};\boldsymbol{\theta_j})+u_j,
\end{align*}
with the location-specific deviation $s(x_{t,j}\ldots x_{t-L,j};\boldsymbol{\theta_j})$ being specified as:
\begin{align*}
    s(x_{t,j}, \ldots , x_{t-L,j} ; \boldsymbol{\theta_{j}})&=\sum_{i=1}^{v_x} \sum_{k=1}^{v_l}\left(\sum_{l=0}^L\widetilde{b}_i(x_{t-l,j})\shortsmile{b}_k(l) \right)\theta_{j,ik}.
\end{align*}
Each $\boldsymbol{\theta_j}$  is a location-specific vector of coefficients $\boldsymbol{\theta_j} = (\theta_{j,11},\ldots, \theta_{j,1v_l}, \ldots, \theta_{j,v_xv_l})^\top$ and the combined parameter vector $\boldsymbol{\theta}_{\mathcal{J}}=(\boldsymbol{\theta_1}^T,\boldsymbol{\theta_2}^T,\ldots,\boldsymbol{\theta_J}^\top)^\top$ has dimension $J v_x v_l$. By specifying the same basis for the common exposure-lag-response curve and its location-specific deviation, the model can be reformulated as:
\begin{align*}
    \log(\mu_{t,j}) = \beta_0+\sum_{h=1}^{H}{\beta_h a_{t,j,h}}+ \sum_{i=1}^{v_x} \sum_{k=1}^{v_l}\left(\sum_{l=0}^L\widetilde{b}_i(x_{t-l,j})\shortsmile{b}_k(l) \right)(\theta_{ik}+\theta_{j,ik})+u_j.
\end{align*}
Through this notation, $\theta_{j,ik}$ is interpreted as a random deviation from the common parameter $\theta_{ik}$. 
Inspired by the four types of spatial-temporal interactions of \cite{Knorr-Held2000}, we define four alternative prior structures on $\boldsymbol{\theta}_{\mathcal{J}}$ (Type I-IV), representing all combinations of (1) the use of penalized (P-splines) versus unpenalized (e.g. natural) splines and (2) the use of an independent versus spatially structured location-specific deviation from the common curve.

\subsubsection*{Prior Type I: Unpenalized and Independent}
Type I prior distributions correspond to unpenalized DLNMs with an exposure-lag-response relationship that is randomly different from region to region, without any spatial structure. Hence, as mentioned earlier, the prior distribution on $\boldsymbol{\theta}$ is simply $\boldsymbol{\theta} \sim \mathcal{N}(0,\zeta^{-1} I)$. Furthermore, if we model $\boldsymbol{\theta}_{\mathcal{J}}$ as an independent random deviation from this common structure, we can assume the following prior for the location-specific parameters:
\begin{align*}
    \boldsymbol{\theta}_{\mathcal{J}} \sim \mathcal{N}\biggl(0,\tau_\theta^{-1} I\biggl), 
\end{align*} with $\tau_\theta$ representing the precision parameter.  This corresponds to an unpenalized, spatially independent DLNM.

\subsubsection*{Prior Type II: Penalized and Independent}
Type II prior distributions correspond to penalized DLNMs with location-specific exposure-lag-response relationships that are still independent in space. Hence, assuming $\boldsymbol{\theta} \sim \mathcal{N}(0,\mathcal{P}(\boldsymbol{\lambda}^{(1)})^{-1})$ with penalty parameters $\boldsymbol{\lambda}^{(1)} = (\lambda_{x_1},\lambda_{l_1})^\top$ that controls smoothness of the common DLNM surface, we adopt the following structure for the location-specific effects $\boldsymbol{\theta}_{\mathcal{J}}$:
\begin{align*}
    \boldsymbol{\theta}_{\mathcal{J}} &\sim \mathcal{N}\biggl(0,(I \otimes  \mathcal{P}(\boldsymbol{\lambda}^{(2)}))^{-1}\biggl),
\end{align*}
with penalty parameters $\boldsymbol{\lambda}^{(2)} = (\lambda_{x_2},\lambda_{l_2})^\top$ that governs the smoothness of the location-specific deviations. This is a penalized, spatially independent DLNM.

\subsubsection*{Prior Type III: Unpenalized and Spatially structured}
We let Type III prior distributions correspond to unpenalized DLNMs, i.e. $\boldsymbol{\theta} \sim \mathcal{N}(0,\zeta^{-1} I)$,  with a spatially structured exposure-lag-response relationship. The prior distribution on $\boldsymbol{\theta}_{\mathcal{J}}$ is thus specified as:
\begin{align*}
    \boldsymbol{\theta}_{\mathcal{J}} &\sim \mathcal{N}\biggl(0,\tau_\theta^{-1}(Z(\rho^{(1)}) \otimes  I)^{-1}\biggl),
\end{align*}
where $Z(\rho)$, with $0 \leq \rho < 1$, is the Leroux precision matrix \citep{leroux1999}, i.e. $Z(\rho) = \rho \Lambda  + (1-\rho)I_J$ and $\Lambda $ has elements
\[ r_{jh} = \begin{cases} 
      n_j & j = h \\
      -1 & j \sim h \\
      0 & \text{otherwise},
   \end{cases}
\]
with $j \sim h$ denoting that location $j$ and $h$ are neighbours. This prior provides a flexible compromise between intrinsic CAR structure ($\rho=1$) and independence ($\rho=0$).
This results in an unpenalized, spatially structured DLNM.

\subsubsection*{Prior Type IV: Penalized and Spatially structured}
Finally, let Type IV priors correspond to penalized DLNMs, i.e. $\boldsymbol{\theta} \sim \mathcal{N}(0,\mathcal{P}(\boldsymbol{\lambda}^{(1)})^{-1})$, with exposure-lag-response relationships that are spatially structured. We therefore specify the prior distribution on $\boldsymbol{\theta}_{\mathcal{J}}$ as:
\begin{align*}
    \boldsymbol{\theta}_{\mathcal{J}} &\sim \mathcal{N}\biggl(0,(Z(\rho^{(1)}) \otimes  \mathcal{P}(\boldsymbol{\lambda}^{(2)}))^{-1}\biggl),
\end{align*}
with $Z(\rho)$ and $\mathcal{P}(\boldsymbol{\lambda})$ as specified before. This is a penalized, spatially structured DLNM.

\subsection{Bayesian model formulation}
Following \cite{Rutten2026}, a Gaussian prior on the fixed effects parameter is assumed. Hence, $\boldsymbol{\beta} = (\beta_0, \beta_1, \ldots, \beta_H)^\top \sim \mathcal{N}(\boldsymbol{0},\Omega^{-1}_\beta)$ with $\Omega_\beta = \zeta I_{H+1}$. The precision matrix for the spatial random effect $\boldsymbol{u} = (u_1, \ldots u_J)^\top$ can be written as $G$. Different structures can be assumed on $G$. When working with a Type I or Type II prior, no spatial structure on the location-specific exposure-lag-response curve is assumed. Therefore, it is also natural to assume $u_j \stackrel{iid}{\sim} \mathcal{N}(0, \tau_u^{-1})$ with precision parameter $\tau_u$ such that $G = \tau_u I_J$. When working with a Type III or Type IV prior, assuming a Leroux spatial structure, it is reasonable to assume a similar structure on the random effects, i.e. $G = \tau_u Z(\rho^{(2)})$. Other (spatial) structures, such as an intrinsic autoregressive (ICAR) prior or BYM structure \citep{besag1991} may also be used. 

Define now $V_\theta$ and $V_{\theta_{\mathcal{J}}}$ as the prior precision matrices of $\boldsymbol{\theta}$ and $\boldsymbol{\theta}_{\mathcal{J}}$, respectively, depending on the chosen structure. The precision matrix of $\boldsymbol{\xi} = (\boldsymbol{\beta}^\top, \boldsymbol{\theta}^\top,\boldsymbol{\theta}_{\mathcal{J}}^\top, \boldsymbol{u}^\top)^\top$ can then be defined as $Q = \text{blkdiag}(\Omega_\beta,V_\theta, V_{\theta_{\mathcal{J}}}, G)$ with $\text{blkdiag}(\cdot)$ denoting a block-diagonal matrix.

Besides, denote $A$ as the $TJ \times (H+1)$ dimensional design matrix, including the intercept, for the $H$ linear covariates. Moreover, denote by $W$ the $TJ \times (v_xv_l)$ dimensional matrix for which the $tj$-th row is equal to $\boldsymbol{w_{t,j}}$, and define $M$ as the $TJ \times J$ dimensional design matrix for the random vector $\boldsymbol{u}$, such that row $tj$ is equal to $1$ in column $j$ and $0$ otherwise. All data can then be combined in the design matrix $X = [A:W:M\otimes_r W:M]$ with row $tj$ of matrix $M\otimes_r W$ equal to $\boldsymbol{m_{t,j}}^\top \otimes \boldsymbol{w_{t,j}}^\top$, where $\boldsymbol{m_{t,j}}^\top$ and $\boldsymbol{w_{t,j}}^\top$ represent row $tj$ of $M$ and $W$, respectively.

Combine all penalty parameters $\lambda$ in the vector $\boldsymbol{\bar{\lambda}}$ and all precision parameters $\tau$ in the vector $\boldsymbol{\bar{\tau}}$. If applicable, denote the collection of correlation parameters by $\boldsymbol{\bar{\rho}}$. Assuming that the counts $\boldsymbol{y} = (y_{1,1}, y_{2,1},\ldots y_{T,J})^\top$ follow a Poisson distribution, we can write the full Bayesian model as:
\begin{align*}
            &(\boldsymbol{y}|\boldsymbol{\xi}) \sim \text{Poisson}(\boldsymbol{\mu}) \text{ with } \log(\boldsymbol{\mu}) = X\boldsymbol{\xi}, \\
            & (\boldsymbol{\xi}|\boldsymbol{\bar{\lambda}}, \boldsymbol{\bar{\tau}}, \boldsymbol{\bar{\rho}}) \sim \mathcal{N}(\boldsymbol{0},Q^{-1}), \\
             & (\tau_i|\delta_{\tau_i}) \sim \mathcal{G}\left(\frac{\nu}{2},\frac{\nu \delta_{\tau_i}}{2}\right) \text{for all $\tau_i \in \boldsymbol{\tau}$} , \\
             & \delta_{\tau_i} \sim \mathcal{G}(a,b), \\
             &(\lambda_i|\delta_{\lambda_i}) \sim \mathcal{G}\left(\frac{\nu}{2},\frac{\nu \delta_{\lambda_i}}{2}\right) \text{for all $\lambda_i \in \boldsymbol{\bar{\lambda}}$} ,\\
             & \delta_{\lambda_i} \sim \mathcal{G}(a,b), \\
             & \rho_i \sim \text{Beta}\left(\frac{1}{2}, \frac{1}{2}\right) \text{for all $\rho_i \in \boldsymbol{\bar{\rho}}$} ,
\end{align*}
with $\mathcal{G}(a,b)$ a Gamma distribution with mean $a/b$ and variance $a/b^2$. The parameters $a$ and $b$ are chosen sufficiently small (e.g. $a=b=10^{-5}$) and $\nu$ is fixed (e.g. $\nu = 3$), resulting in a robust prior specification \citep{jullion2007}. 

Note that the model can easily be extended towards a negative binomial model, i.e. $(\boldsymbol{y}|\boldsymbol{\xi}) \sim \text{NegBin}(\boldsymbol{\mu}, \phi)$, with $\log(\boldsymbol{\mu}) = X\boldsymbol{\xi}$ and $\phi \propto \phi^{-1}$ the dispersion parameter (i.e. the variance of $y_{t,j}$ is equal to $\mu_{t,j}+\frac{\mu_{t,j}^2}{\phi}$).

\subsection{Laplace approximation}
To efficiently estimate the parameters $\boldsymbol{\xi}$ in this complex model, we use Laplace approximations. This method constructs a Gaussian approximation to the posterior distribution, yielding significant computational gains compared to the traditional Markov chain Monte Carlo (MCMC) methods \citep{gressani2018, lambert2023, Rutten2026}. 

In the first step, the conditional posterior of $\boldsymbol{\xi}$, given the hyperparameters, can be written as $$p(\boldsymbol{\xi}| \boldsymbol{\bar{\lambda}}, \boldsymbol{\bar{\tau}}, \boldsymbol{\bar{\rho}},\phi; \mathcal{D}) \propto \mathcal{L}(\boldsymbol{\xi},\boldsymbol{\bar{\lambda}}, \boldsymbol{\bar{\tau}}, \boldsymbol{\bar{\rho}},\phi; \mathcal{D})p(\boldsymbol{\xi}| \boldsymbol{\bar{\lambda}}, \boldsymbol{\bar{\tau}}, \boldsymbol{\bar{\rho}},\phi) $$with $$p(\boldsymbol{\xi}| \boldsymbol{\bar{\lambda}}, \boldsymbol{\bar{\tau}}, \boldsymbol{\bar{\rho}},\phi)  \propto \exp\bigg( - \frac{1}{2}{\boldsymbol{\xi}}^{\top} Q \boldsymbol{\xi}\bigg)$$ and $\mathcal{L}(\boldsymbol{\xi},\boldsymbol{\bar{\lambda}}, \boldsymbol{\bar{\tau}}, \boldsymbol{\bar{\rho}},\phi; \mathcal{D})$ the Poisson or negative binomial likelihood. After analytically deriving the gradient and Hessian of this conditional posterior with respect to $\boldsymbol{\xi}$, the posterior mode $\hat{\boldsymbol{\xi}}$ can be estimated using the Newton-Raphson algorithm. This yields a multivariate Gaussian approximation to $p(\boldsymbol{\xi}| \boldsymbol{\bar{\lambda}}, \boldsymbol{\bar{\tau}}, \boldsymbol{\bar{\rho}},\phi; \mathcal{D})$ given by $$\widetilde{p}_G(\boldsymbol{\xi}| \boldsymbol{\bar{\lambda}}, \boldsymbol{\bar{\tau}}, \boldsymbol{\bar{\rho}},\phi; \mathcal{D}) = \mathcal{N}(\hat{\boldsymbol{\xi}},\hat{\Sigma}),$$ where $\hat{\Sigma}$ is the inverse of the negative Hessian matrix evaluated in $\hat{\boldsymbol{\xi}}$. 

In a second step, the hyperparameters can be optimized following the approach of \cite{Rutten2026}, obtaining estimates that are used to derive the marginal posterior of $\boldsymbol{\xi}$. More details can be found in the Supplementary Materials.

\subsection{Model comparison}
Different model comparison criteria are implemented. Following \cite{Spiegelhalter2002}, the Deviance Information Criteria (DIC) is defined as $$D(\hat{\boldsymbol{\xi}}_\lambda)+2p_D$$ with $$D(\boldsymbol{\xi}) = -2\log(\mathcal{L}(\boldsymbol{\xi},\boldsymbol{\bar{\lambda}}, \boldsymbol{\bar{\tau}}, \boldsymbol{\bar{\rho}},\phi; \mathcal{D})).$$ In this notation, $p_D$ represents the effective number of parameters, approximated by $p_D \approx n_{\boldsymbol{\xi}}-\text{Trace}(Q\hat{\boldsymbol{\Sigma}}_\lambda)$, with $n_{\boldsymbol{\xi}}$ the number of parameters in $\boldsymbol{\xi}$ and $ \hat{\boldsymbol{\Sigma}}_\lambda$ and $Q$ the posterior covariance and prior precision matrix of $\boldsymbol{\xi}$ respectively \citep{rue2009}. 

Watanabe-Akaike Information Criterion / Widely Applicable Information Criterion (WAIC) \citep{Watanabe2010} is calculated as $$-2(\text{lppd}-p_{\text{WAIC}}),$$ with lppd the log posterior predictive density (lppd) and the effective number of parameters $p_{\text{WAIC}}$ defined using equation (11) of \cite{Gelman2014}.

Lastly, we implemented the Conditional Predictive Ordinate (CPO) \citep{Congdon2020}. The CPO for a given data point can be calculated as $$CPO_i = p(y_i \vert \mathcal{D}_{(i)}) = \biggl(E_{p(\boldsymbol{\xi}|  \mathcal{D})}\bigl(\frac{1}{p(y_{i}\vert\boldsymbol{\xi})}\bigl)\biggl)^{-1},$$ where $\mathcal{D}_{(i)}$ represents all data except for observation $y_i$. These CPO values are often summarized by $-\sum_{i}{\log(CPO_i)}$ such that a lower value indicates a better model fit, similar to DIC and WAIC.

\section{Simulation study}
A simulation study has been conducted to evaluate the model performance under various settings. Settings differ in population size, strength of the spatial correlation and heterogeneity of the effects. 
\subsection{Simulation set-up}
\subsubsection*{Simulating data}
The time-varying predictor $x_t$ is the daily summer temperature (i.e. June to September), during the period 2007-2016 in Barcelona, standardized over a range $0-10$. Temperature data are available in the $73$ neighbourhoods of Barcelona separately, resulting in $1220$ observations ($t = 1,\ldots,1220$) for each neighbourhood \citep{Quijal-Zamorano2024}. A neighbourhood-specific count time series $y_{t,j}$ was simulated from a Poisson distribution with mean:
\begin{align*}
    \log(\mu_{t,j}) = \beta_0 + \sum_{l=0}^{8}{f_j \cdot w_j(x_{t-l,j},l)+u_j+\log(\text{pop\_size}_j})
\end{align*}
with $u_j$ an area-specific random effect, simulated from a Leroux distribution with $\rho = 0.90$ and $\sigma^2 = 0.05$. The intercept $\beta_0 = 10.5$ is chosen such that the daily average response count represents a realistic European mortality rate of $0.03$ daily deaths per $1000$ inhabitants on average.

\begin{figure}[h]
    \centering
    \includegraphics[width=1\linewidth]{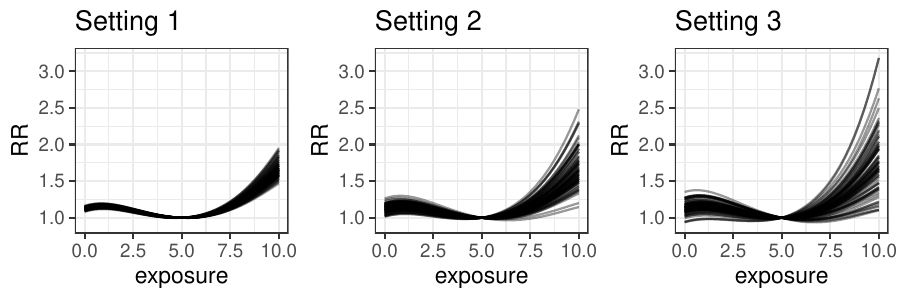}
    \caption{Assumed exposure-response relationship (overall cumulative RR) for different regions of Barcelona (in the three different settings).}
    \label{fig:simulation_scenario}
\end{figure}

\begin{figure}[h]
    \centering
    \includegraphics[width=\linewidth]{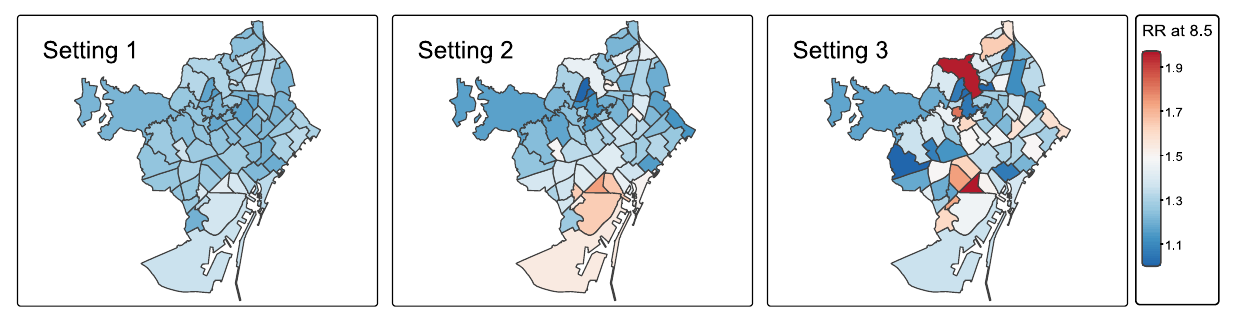}
    \caption{Assumed overall cumulative RR at exposure level 8.5 on a map of Barcelona (in the three different settings).}
    \label{fig:simulation_scenario_map}
\end{figure}

The function $f_j \cdot w_j(x_{t-l,j},l)$, describing the assumed location-specific exposure-lag-response curve, depends on the simulation setting. Three different settings were considered: (1) strong spatial correlation with small heterogeneity of the effect, (2) strong spatial correlation with large heterogeneity of the effect and (3) small spatial correlation with large heterogeneity of the effect. A figure with the overall cumulative (over lag $0-8$) exposure-response curve for every location and a map of the overall cumulative RR at $8.5$ can be found in Figure \ref{fig:simulation_scenario} and Figure \ref{fig:simulation_scenario_map}, respectively. The 3D exposure-lag-response curve for three different areas, in each of the three settings, is given in the Supplementary Materials (Figure S1).

Each of these settings is considered in a small-area setting, with population sizes resulting from a lognormal distribution with mean on the log-scale equal to $\log(12000)$ and $\sigma = 1.8$ (comparable to the population sizes of Belgian municipalities), and a large-area setting, with population sizes simulated from a lognormal distribution with mean on the log-scale equal to $\log(6000000)$ and $\sigma = 0.9$ (comparable to the population sizes of U.S. states).

\subsubsection*{Fitted models}
In each setting, we fitted spatially-varying DLNMs with priors of Type~I to IV. The penalized models (Type~II and Type~IV) were defined using P-splines with $6$ degrees of freedom in both the exposure and the lag dimension of the DLNM cross-basis, before imposing constraints. An additional penalty matrix was imposed in the lag dimension to encourage shrinkage at longer lags. This penalty matrix was defined as $S_{l_2} = P_{v_l}+\delta I_{v_l}$ with $P_{v_l}$ a diagonal matrix with weights $p_k = k^2$ for $k = 0,\ldots (v_l-1)$. For the unpenalized models (Type~I and Type~III), natural splines with $2$ knots in the exposure dimension (on the $10\%$ and $90\%$ percentile) and $2$ knots in the lag dimension, equally spaced on the log-scale, were defined. We compared the performance of these four models with the common DLNM-LPS model, without spatially varying exposure-lag-response curves, defined in \cite{Rutten2026} and explained in Section~\ref{sec:common}. Furthermore, the methods are compared to the meta-analysis method, proposed by \cite{Gasparrini2013}. In the first stage of this method, a location-specific DLNM is fitted, resulting in location-specific parameters. These location-specific parameters $\boldsymbol{\hat{\theta}_j}$ are pooled, using: 
\begin{align*}
    \boldsymbol{\hat{\theta}_j} \sim \mathcal{N}(\boldsymbol{\vartheta},S_j+\Psi), 
\end{align*}
with 
$\boldsymbol{\vartheta}$ the pooled estimate, $S_j$ the variance-covariance matrix of $\boldsymbol{\hat{\theta}_j}$, obtained from the first stage models, and $\Psi$ representing the between-location variance-covariance matrix, reflecting the between-area variability that cannot be explained by within-location sampling error $S_j$. In this simulation study, similar to Type~I and Type~III models, natural splines with $2$ knots in the exposure dimension and lag dimension are used to define the cross-basis of this method.

\subsubsection*{Model performance}
In every neighbourhood of Barcelona $j = 1,\ldots,73$, we calculated the lag-specific and overall cumulative (over lag $0-8$) log relative risk (RR) by:
\begin{align*}
   \log\bigl(RR_{x,x_0,j}(l)\bigl) &=\sum_{i=1}^{v_x}\sum_{k=1}^{v_l}{\bigl(\widetilde{b}_i(x)-\widetilde{b}_i(x_0)\bigl)\shortsmile{b}_k(l)(\hat{\theta}_{ik}+\hat{\theta}_{j,ik})}, 
\end{align*}
and
\begin{align*}
      \log\bigl(RR_{x,x_0,j}^{\text{overall}}\bigl) &=\sum_{i=1}^{v_x}\sum_{k=1}^{v_l}\biggl(\sum_{l=0}^{L}{\bigl(\widetilde{b}_i(x)-\widetilde{b}_i(x_0)\bigl)\shortsmile{b}_k(l)\biggl)(\hat{\theta}_{ik}+\hat{\theta}_{j,ik})}, 
\end{align*}
for a grid of exposure levels $(x=0,0.25,\ldots,9.75,10)$, compared to a reference exposure level $x_0 = 5$. These metrics were summarized using the root mean squared error (RMSE) and coverage (see Supplementary Materials for more detailed information). Moreover, we calculated the AUC (area under the receiver operating characteristic (ROC) curve) to evaluate the discriminating power of the model \citep{bamber1975}. More specifically, we defined, for each exposure level, which areas correspond to the top $25\%$ highest risk areas, based on the true RRs, resulting in a binary outcome (high/low risk) for each region. The AUC of the ROC curve then illustrates how well the estimated area-specific RRs can discriminate between these true high and low risk areas. A similar metric was defined for the top $10\%$ highest risk areas. An AUC of $0.5$ would reflect no discriminating power, while an AUC of $1$ would indicate perfect discrimination. We summarized these metrics by taking the average AUC over the grid of exposure levels.

\subsection{Simulation results}
We performed $250$ simulation runs in every simulation setting, using the Flemish Super Computer (VSC). The results can be found in Table \ref{tab:simulation_study_small} and \ref{tab:simulation_study_large} for the small and large area case, respectively. Moreover, the estimated overall-cumulative RR, averaged over $250$ runs, for every neighbourhood of Barcelona at an exposure level of $8.5$ is included in Figure \ref{fig:setting2_map} for Setting~2, and in the Supplementary Materials for Setting~1 and Setting~3 (Figure S2 and Figure S3 respectively).

\begin{table}[h]
\caption{Simulation results for small area setting: RMSE of lag-specific RR (RMSE RR), RMSE of cumulative RR, over lag $0-8$, (RMSE RR overall), coverage of lag-specific RR (cov RR),  coverage of cumulative RR (cov RR overall), AUC of top $10\%$ and AUC of top $25\%$ high risk areas. The computation time is reported in seconds.}
\label{tab:simulation_study_small}
\centering
\begin{adjustbox}{width=\linewidth} % Optional for scaling if necessary
\begin{tabular}{llccccccc}
\hline
 \textbf{Setting} &  \textbf{Method} & \textbf{time} & \textbf{cov RR} &\textbf{cov RR} & \textbf{RMSE RR} &\textbf{RMSE RR} & \textbf{AUC $10\%$} & \textbf{AUC $25\%$} \\ 
  &   &  &  & \textbf{overall} &  &\textbf{overall} & & \\ \hline
   \multirow{6}{*}{Setting 1} & Type I  & $25.34$& $0.9962$& $0.9933$& $0.0150$ & $0.0697$ & $0.6175$& $0.5771$\\
                          & Type II  & $59.92$ & $0.9984$ & $0.9951$ & $0.0118$ & $0.0694$ & $0.6236$&  $0.5822$\\
                          & Type III  & $29.96$ & $0.9917$ & $0.9906$ & $0.0141$& $0.0602$ & $0.6527$ & $0.6067$ \\
                          & Type IV  & $124.36$ & $0.9969$ &  $0.9946$& $0.0104$& $0.0574$ & $0.6624$& $0.6125$ \\
                          & Common  & $27.25$ & $0.9467$& $0.8208$ & $0.0094$& $0.0503$ & $0.5000$ & $0.5000$ \\
                          & Meta  & $25.62$ & $0.9831$ & $0.9668$& $0.0266$& $0.1063$ & $0.5982$ & $0.5590$\\

\hline
 \multirow{6}{*}{Setting 2} & Type I  & $24.68$ & $0.9945$ & $0.9885$ & $0.0165$ & $0.0856$& $0.6815$ & $0.6574$\\
                          & Type II  & $59.17$ & $0.9979$ & $0.9920$ & $0.0129$& $0.0795$ & $0.7089$& $0.6852$ \\
                          & Type III  & $29.06$ & $0.9860$ & $0.9757$ & $0.0156$& $0.0776$ & $0.7750$&$0.7196$ \\
                          & Type IV  & $121.14$ & $0.9952$ & $0.9862$& $0.0112$ & $0.0689$ & $0.8059$& $0.7528$ \\
                          & Common  & $26.45$ & $0.8392$ & $0.4944$ & $0.0124$& $0.0792$ & $0.5000$ & $0.5000$\\
                          & Meta  & $24.41$ & $0.9767$ & $0.9475$ & $0.0279$& $0.1215$ & $0.6638$ & $0.6414$\\

\hline
 \multirow{6}{*}{Setting 3} & Type I  & $25.79$ & $0.9845$& $0.9707$& $0.0189$ & $0.1095$ & $0.7618$ & $0.7134$ \\
                          & Type II  & $61.55$ & $0.9951$& $0.9807$& $0.0151$ & $0.1008$ & $0.8101$& $0.7539$  \\
                          & Type III  & $30.09$& $0.9678$& $0.9203$ & $0.0185$ & $0.1083$ & $0.7709$ & $0.7122$ \\
                          & Type IV  & $125.05$ & $0.9807$ & $0.9470$ & $0.0148$& $0.1001$& $0.8182$ & $0.7521$ \\
                          & Common  & $27.51$ & $0.7511$ & $0.3774$& $0.0165$& $0.1164$& $0.5000$ & $0.5000$\\
                          & Meta  & $25.14$ & $0.9702$ & $0.9410$ & $0.0291$ & $0.1333$& $0.7805$&$0.7245$\\

\hline

\end{tabular}
\end{adjustbox}
\end{table}

\begin{table}[h]
\caption{Simulation results for large area setting: RMSE of lag-specific RR (RMSE RR), RMSE of cumulative RR, over lag $0-8$, (RMSE RR overall), coverage of lag-specific RR (cov RR),  coverage of cumulative RR (cov RR overall), AUC of top $10\%$ and AUC of top $25\%$ high risk areas. The computation time is reported in seconds.}
\label{tab:simulation_study_large}
\centering
\begin{adjustbox}{width=\linewidth} % Optional for scaling if necessary
\begin{tabular}{llccccccc}
\hline
 \textbf{Setting} &  \textbf{Method} & \textbf{time} & \textbf{cov RR} &\textbf{cov RR} & \textbf{RMSE RR} &\textbf{RMSE RR} & \textbf{AUC $10\%$} & \textbf{AUC $25\%$} \\ 
  &   &  &  & \textbf{overall} &  &\textbf{overall}  \\ \hline
 \multirow{6}{*}{Setting 1} & Type I  & $21.88$& $0.9308$ & $0.8718$& $0.0067$ & $0.0385$ & $0.8695$& $0.8367$ \\
                          & Type II  & $55.81$& $0.9823$& $0.9636$&$0.0046$ & $0.0235$& $0.8830$& $0.8539$  \\
                          & Type III & $25.48$ & $0.9183$& $0.8513$ & $0.0060$&$0.0367$ & $0.8834$ & $0.8472$ \\
                          & Type IV  & $118.17$& $0.9841$ &$0.9652$ & $0.0037$& $0.0201$& $0.9015$ &  $0.8722$ \\
                          & Common  & $23.89$& $0.5883$&$0.2566$ & $0.0038$&$0.0273$ & $0.5000$& $0.5000$ \\
                          & Meta  & $28.34$& $0.8350$&$0.7985$ & $0.0055$& $0.0340$&$0.9326$ & $0.9008$\\

\hline
 \multirow{6}{*}{Setting 2} & Type I  & $21.57$& $0.9210$&$0.8455$ & $0.0077$& $0.0443$& $0.9490$& $0.9441$\\
                          & Type II  & $55.38$ &$0.9752$ & $0.9521$& $0.0050$& $0.0267$&$0.9677$ & $0.9671$ \\
                          & Type III  & $24.93$&$0.9083$ & $0.8212$&$0.0070$ &$0.0425$ &$0.9601$ &$0.9511$ \\
                          & Type IV  & $116.13$& $0.9735$& $0.9472$&$0.0043$ & $0.0241$& $0.9752$& $0.9739$ \\
                          & Common  & $23.59$&$0.3348$ & $0.0826$& $0.0095$&$0.0696$ & $0.5000$& $0.5000$\\
                          & Meta  & $27.87$ & $0.8441$&$0.8531$ &$0.0055$ & $0.0341$& $0.9894$&$0.9885$\\

\hline
 \multirow{6}{*}{Setting 3} & Type I  & $22.60$& $0.9072$& $0.8221$& $0.0085$& $0.0482$& $0.9769$& $0.9635$\\
                          & Type II  & $57.71$& $0.9638$ & $0.9315$& $0.0055$& $0.0307$&$0.9882$ & $0.9779$ \\
                          & Type III  &$26.30$ & $0.8878$ &$0.7747$ & $0.0085$&$0.0514$ &$0.9719$ & $0.9567$\\
                          & Type IV  &$120.60$ & $0.9507$& $0.9122$& $0.0052$&$0.0317$ & $0.9905$&  $0.9801$\\
                          & Common  & $24.75$&$0.2869$ &$0.0761$ & $0.0140$& $0.1068$&$0.5000$ &$0.5000$ \\
                          & Meta  & $28.23$&$0.8381$ &$0.8697$ & $0.0057$&$0.0349$ & $0.9956$&$0.9904$\\

\hline

\end{tabular}
\end{adjustbox}
\end{table}

\begin{table}[h]
\caption{Percentage of the simulation runs each method is preferred according to DIC, WAIC and CPO. Results are given for both small and large areas and in each simulation setting.}
\label{tab:model_selection}
\centering
\begin{tabular}{ll|ccc|ccc}
\hline
& & \multicolumn{3}{c|}{Small area} & \multicolumn{3}{c}{Large area} \\
 \textbf{Setting} &  \textbf{Method} & \textbf{DIC} & \textbf{WAIC} &\textbf{CPO} & \textbf{DIC} & \textbf{WAIC} &\textbf{CPO}\\ 
 \hline
  \multirow{5}{*}{Setting 1} & Type I  & $0.00$ & $0.00$ & $0.04$ & $0.00$&$0.00$ & $0.00$\\
                          & Type II  &$0.00$  & $0.00$ & $0.10$ &$0.00$ & $0.00$&$0.00$ \\
                          & Type III  & $0.02$ & $0.02$ & $0.04$& $0.00$& $0.00$&$0.00$\\
                          & Type IV  & $0.04$ & $0.04$ & $0.10$& $1.00$&$1.00$ &$1.00$\\
                          & Common  & $0.94$ & $0.94$ & $0.72$ &$0.00$ &$0.00$ &$0.00$\\
                          \hline
                         \multirow{5}{*}{Setting 2} & Type I  &$0.00$ & $0.00$&$0.00$ &$0.00$ & $0.00$&$0.00$\\
                          & Type II  & $0.00$ &$0.00$ &$0.00$&$0.00$ &$0.00$ &$0.00$ \\
                         & Type III  & $0.14$ &$0.14$ &$0.14$& $0.00$&$0.00$ &$0.00$\\
                          & Type IV  &$0.74$ &$0.75$ &$0.75$ &$1.00$ &$1.00$ &$1.00$\\
                          & Common  &$0.12$ &$0.11$ &$0.11$&$0.00$ &$0.00$ &$0.00$\\
                          \hline
                         \multirow{5}{*}{Setting 3} & Type I  &$0.00$ & $0.00$&$0.00$ &$0.00$ & $0.00$&$0.00$\\
                          & Type II  & $0.01$ &$0.01$ &$0.00$&$0.00$ &$0.00$ &$0.00$ \\
                         & Type III  & $0.02$ &$0.03$ &$0.03$& $0.00$&$0.00$ &$0.00$\\
                          & Type IV  &$0.97$ &$0.96$ &$0.97$ &$1.00$ &$1.00$ &$1.00$\\
                          & Common  &$0.00$ &$0.00$ &$0.00$&$0.00$ &$0.00$ &$0.00$\\
 
\hline

\end{tabular}
\end{table}

\begin{figure}[h]
    \centering
    \begin{subfigure}{\textwidth}
        \centering
        \includegraphics[width=\linewidth]{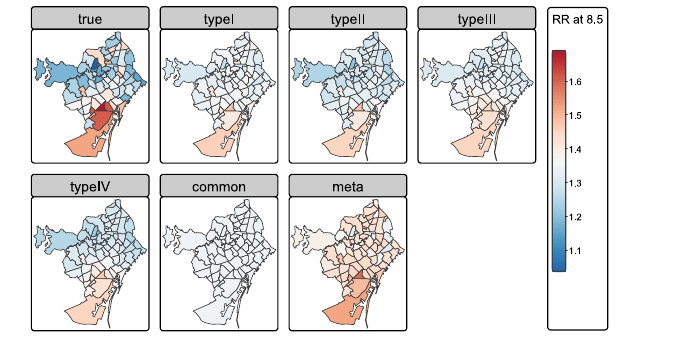}
         \caption{}
        \label{fig:setting2_map_small}
    \end{subfigure}
    \hfill
    \begin{subfigure}{\textwidth}
        \centering
        \includegraphics[width=\linewidth]{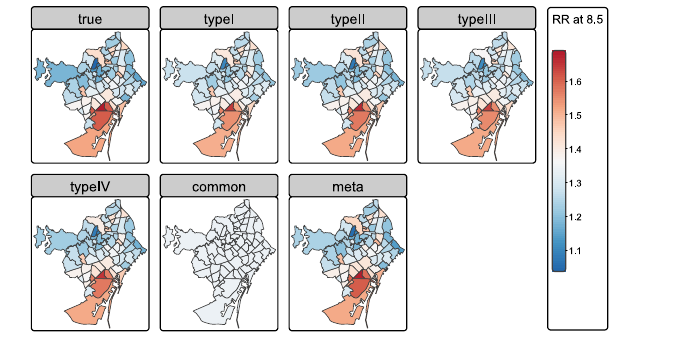}
        \caption{}
        \label{fig:setting2_map_large}
    \end{subfigure}
    \caption{Estimated overall cumulative RR for Setting~2 (averaged over the $250$ simulation runs) in the small area scenario (a) and large area scenario (b).}
    \label{fig:setting2_map}
\end{figure}

The results in Table \ref{tab:simulation_study_small} show that, in small area settings, all proposed one-stage spatial DLNM-LPS models outperform the two-stage meta-analysis method in terms of RMSE. Although the assumption of a common exposure-lag-response curve shows similar RMSEs compared to the spatial models, this assumption results in undercoverage of the credible intervals in the two settings with large spatial heterogeneity (Setting~2 and Setting~3). In contrast, satisfying coverage can be seen for all one-stage spatial models and for the two-stage meta-analysis, in all three settings. By definition, the common exposure-lag-response model naturally has no discriminating power between high- and low-risk areas ($\text{AUC}=0.5$). Moreover, we see that for all other methods, discrimination is more difficult in Setting~1, compared to the other two settings. This is not surprising since the spatial heterogeneity in Setting~1 is low. Table \ref{tab:model_selection}, showing the percentage of runs a certain method is preferred according to model selection criteria, also confirms that the common model is often preferred in Setting~1. In Setting~2, the AUC shows acceptable discrimination for the proposed one-stage methods, especially for the two spatial methods (Type~III and Type~IV). Hence, although there is little gain in RMSE by assuming spatially-varying exposure-lag-response curves, additional insights about low and high risk areas can be obtained, compared to a common curve. This is confirmed by Figure \ref{fig:setting2_map_small}, showing discrimination between true low and high risk areas. Table \ref{tab:model_selection} shows that the Type~IV model is preferred in about $75\%$ of the simulation runs, according to all selection criteria. In Setting~3, acceptable discrimination is obtained for all four proposed one-stage methods. Differences in AUC between the spatially-structured (Type~III and Type~IV) and spatially-unstructured (Type~I and Type~II) methods are smaller, compared to Setting~2. This observation is not surprising, since Setting~2 reflects a scenario with strong spatial correlation while Setting~3 hardly shows any spatial structure (see Figure \ref{fig:simulation_scenario_map}). Looking at Table~3, the Type~IV model is almost always preferred. Finally, Table \ref{tab:simulation_study_small} shows that the RMSE is always lower for the penalized one-stage methods (Type~II and Type~IV) compared to the unpenalized methods (Type~I and Type~III), illustrating the benefit of penalization.

Looking at the results in a large area setting (Table \ref{tab:simulation_study_large}), we observe that the undercoverage of the common model is even more severe. Moreover, in Setting~2 and Setting~3, the spatial models now clearly outperform the common model in terms of RMSE. Figure \ref{fig:setting2_map_large} also confirms good estimation of the region-specific RRs. The two-stage meta-analysis shows similar RMSEs compared to the one-stage spatial methods. Similar to the small area setting, the penalized methods (Type~II and Type~IV) show slightly lower RMSEs compared to the unpenalized methods (Type~I and Type~III). Furthermore, the coverage of the penalized methods is now clearly higher. However, AUC indicates very good discrimination for all spatial models in all settings. Table \ref{tab:model_selection} shows that model selection criteria consistently point towards the Type~IV method. 

In terms of computation time, we can see that the unpenalized (Type~I and Type~III) methods show similar computation times compared to a meta-analysis, but in small area settings, the former methods show remarkably better performance. The computation time of the penalized methods is slightly higher, but as they can still be fitted within a few minutes, the computational burden remains low.

We performed the same simulation study in a negative binomial setting, obtaining similar results (see Supplementary Materials Table S1 and S2)

\section{Data application}
We apply our proposed method to analyse the city-specific relationship between daily mean temperature and mortality in all $391$ municipalities of Sicily, Italy. A dataset containing daily mean temperature in every municipality from $2011$ to $2021$, was extracted from the fifth generation of European Reanalysis (ERA5) dataset \citep{ERA5LandDailyStats2025}. Temperature data were provided at $2$ m above the land surface on a grid of approximately $0.25^{\circ}$, and summarized for each region by averaging all grid cells with a centroid contained within the city boundaries. These city boundaries, as they were defined in $2021$, were downloaded from the Italian National Institute of Statistics (ISTAT) \citep{ISTAT_AdminBoundaries2018}. Moreover, city-specific daily mortality data could also be extracted from a dataset that has been made publicly available by ISTAT \citep{Istat_DecessiCauseMorte_2022}.

A basic DLNM-LPS model with a common exposure-lag-response relationship has been fitted to the data, using P-splines with $6$ degrees of freedom in both the exposure and lag dimension, before imposing constraints. Similar to the simulation set-up, an additional penalty was imposed in the lag dimension, to encourage the effect to shrink at longer lags. A maximum lag period of $8$ days was defined. Besides, the four different types of proposed spatial models were fitted, using P-splines with $6$ degrees of freedom for the Type~II and Type~IV model. The unpenalized models, i.e. Type~I and Type~III, were fitted using natural splines with $2$ knots in the variable dimension, on the $10\%$ and $90\%$ percentile of the daily temperature, and $2$ knots in the lag dimension, equally spaced on the log-scale. In all models, day of the week is included as a categorical variable and a smooth time trend is modelled using natural splines with $7$ degrees of freedom per year. The city-specific population size is included as an offset in the model. We fitted all models using both a Poisson and negative binomial likelihood. The computation times can be found in the Supplementary Materials (Table S4).

\begin{figure}
    \centering
    \includegraphics[width=0.75\linewidth]{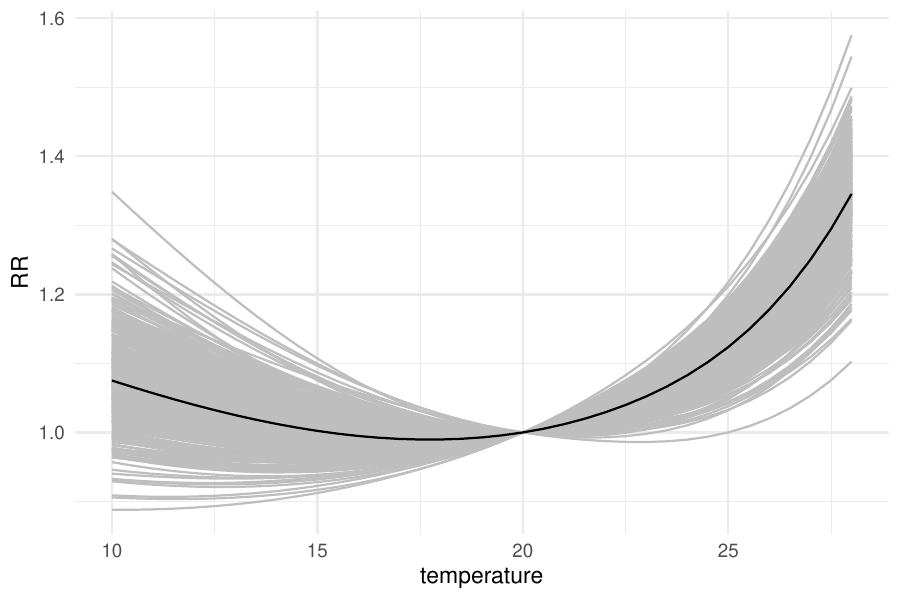}
    \caption{Estimated overall cumulative RR for different temperature values, compared to a reference of $20$ degrees, for all municipalities.}
    \label{fig:overall_sicily}
\end{figure}

\begin{figure}
    \centering
    \includegraphics[width=0.75\linewidth]{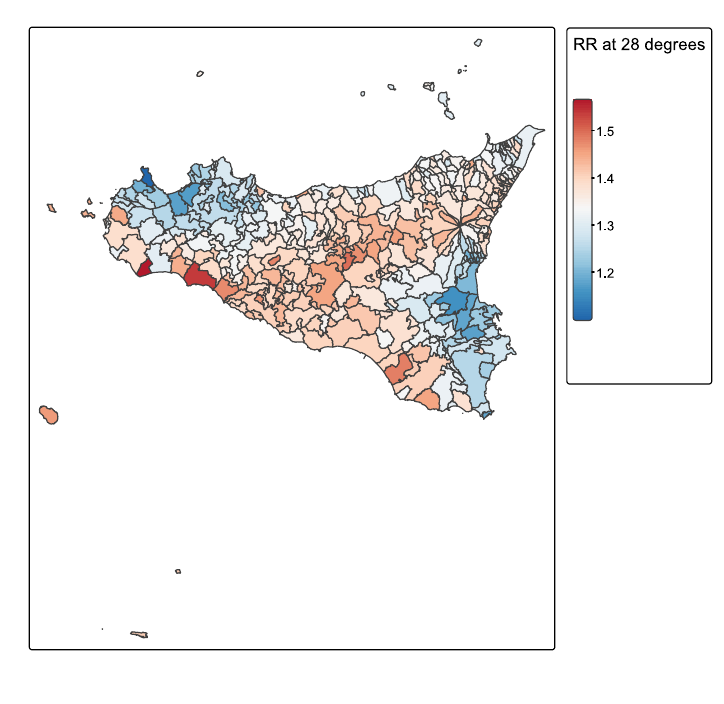}
    \caption{Estimated overall cumulative RR at $28$ degrees, compared to a reference of $20$ degrees, for every municipality in Sicily.}
    \label{fig:map_sicily_28}
\end{figure}

\begin{figure}
    \centering
    \includegraphics[width=0.75\linewidth]{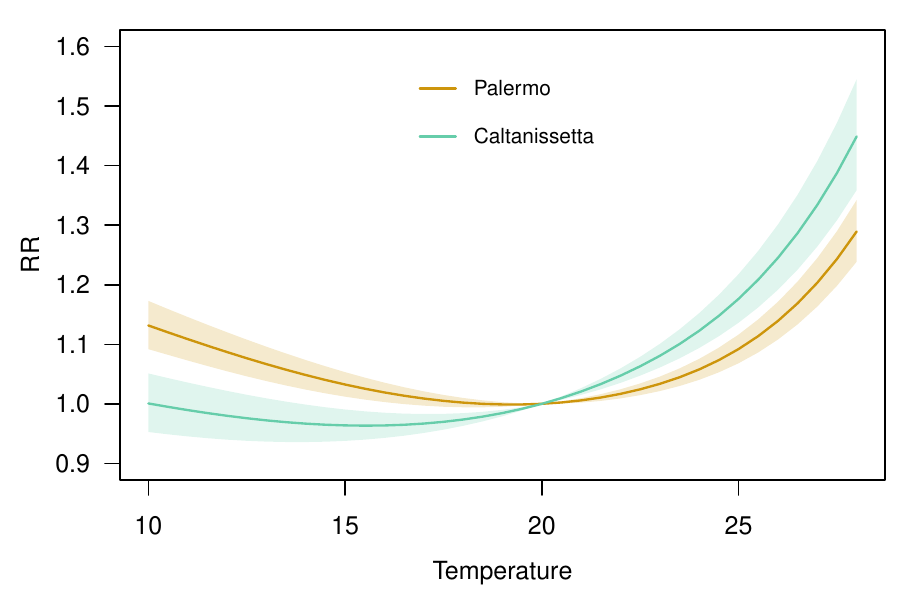}
    \caption{Estimated overall cumulative RR for different temperature values, compared to a reference of $20$ degrees, for Palermo and Caltanissetta, together with the $95\%$ credible interval.}
    \label{fig:overall_RR_sicily}
\end{figure}

All models were compared in terms of DIC, WAIC and CPO ($-\sum_{i}{\log(CPO_i)}$), reporting $\Delta\text{DIC},\Delta\text{WAIC} $ and $\Delta\text{CPO} $, representing the difference between the model specific criteria and the minimum over all models (Table \ref{tab:DIC_sicily}). It can be seen that the Type~IV Poisson model is preferred according to all modelling criteria. Hence, model comparison suggests that there is spatial heterogeneity in the exposure-lag-response relationship between the different municipalities of Sicily. Using this model, Figure \ref{fig:overall_sicily}, shows the overall cumulative RR (over day $0-8$), compared to a reference value of $20$ degrees, for every municipality in Sicily. Moreover, Figure \ref{fig:map_sicily_28}, shows the overall cumulative RR at $28$ degrees on a map, illustrating relatively strong spatial correlation. Figure \ref{fig:overall_RR_sicily} shows significant differences in the estimated overall cumulative RR of Palermo (located in the North-West of Sicily) and Caltanisetta (located in the centre of Sicily). A figure showing the lag-specific RR for these two municipalities at $28$ degrees can be found in the Supplementary Materials (Figure S4). 

\begin{table}
    \centering
        \caption{$\Delta\text{DIC}$, $\Delta\text{WAIC}$ and $\Delta\text{CPO}$, values of the different models, for the Poisson and negative binomial model.}
    \begin{adjustbox}{width=\textwidth}
    \begin{tabular}{r|rrrrr|rrrrr}
        \hline 
        &  \multicolumn{5}{|c|}{Poisson}& \multicolumn{5}{c}{negative binomial} \\
        \hline
        & \multicolumn{1}{|c}{Type I} & Type II & Type III & Type IV & \multicolumn{1}{c|}{Common} & Type I & Type II & Type III & Type IV & Common \\
        \hline
        $\Delta$DIC & $131$ & $94$ & $36$ & $0$& $332$ & $143$ & $443$& $314$ & $494$ & $227$  \\
        \hline
        $\Delta$WAIC & $144$& $97$ & $36$ & $0$& $329$ & $143$& $477$& $323$ & $521$ & $224$ \\
        \hline
        $\Delta$CPO & $63$& $48$ & $17$ & $0$& $164$ & $73$ & $231$ & $158$& $250$& $108$  \\
        \hline
    \end{tabular}
    \end{adjustbox}
    \label{tab:DIC_sicily}
\end{table}

\begin{figure}
    \centering
    \includegraphics[width=\linewidth]{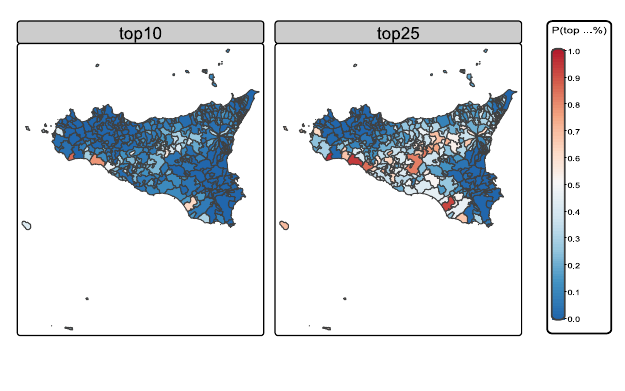}
    \caption{Posterior probability that a certain region belongs to the top $10\%$ and top $25\%$ highest risk areas at $28$ degrees.}
    \label{fig:high_risk_sicily}
\end{figure}

Using the obtained posterior distribution, we can calculate the probability that a certain municipality belongs to the top $10\%$ or top $25\%$ highest risk areas at a temperature of $28$ degrees. A map of these posterior probabilities can be found in Figure \ref{fig:high_risk_sicily}. This figure clearly shows that some areas can be marked as top $25\%$ with a high posterior probability. This again illustrates spatial differences between the regions.
Lastly, we calculated the attributable fraction (AF), introduced in a DLNM context by \cite{Gasparrini2014}, representing the region-specific risk resulting from the temperature exposure in $2021$ (backward perspective). A figure of this AF and the calculated uncertainty, can be found in the Supplementary Materials (Figure S5 and Figure S6). Again, spatial differences can be observed.

\section{Conclusion}
Regions often differ in their socio-demographic, economic, environmental and infrastructural characteristics. When modelling the (possibly delayed) effect of an exposure, these differences likely translate into region-specific exposure-lag-response relationships. However, despite the popularity of DLNMs, a one-stage method that can efficiently model these region-specific effects, was still lacking. Therefore, we introduced an extension of the method of \cite{Rutten2026} towards spatially-varying DLNMs. To efficiently model these effects, we defined four types of priors, differing in spatial structure and penalization.

The simulation study showed that our proposed extension outperforms a two-stage meta-analysis in small area data analyses. This is not surprising since sparse data can result in unstable estimates, making the exchange of information across neighbouring regions crucial for reliable inference. Moreover, we found that our methods show good discriminating high-risk vs low-risk performance in almost every simulation setting. However, when spatial heterogeneity is very low, small area data analyses might not provide sufficient information for the methods to separate high-risk from low-risk areas (Setting~1). In these cases, model comparison showed a preference for a common exposure-lag-response relationship. Moreover, the simulation study showed a preference for penalized methods in terms of RMSE and coverage. As trade off, fitting a penalized method increases computation time. However, it can be seen that all four methods are highly computationally efficient, compared to earlier proposed methods. Although an extensive comparison between our methods and previously proposed methods is not possible due to the computational burden, we provided a small comparison with the \texttt{bam} function in the \texttt{mgcv} package \citep{Economou2025} and the INLA version of the \cite{Quijal-Zamorano2024} method \citep{Quijal-Zamorano2025} in the Supplementary Materials (Table S3). Compared to these methods, our method shows large computational gains indeed, while retaining or even increasing accuracy.

Hence, by introducing a computationally efficient one-stage method that allows for penalization and information sharing, our method shows several advantages over earlier proposed methods. Nevertheless, there are some limitations to our approach. First of all, increasing the degrees of freedom for the penalized models (Type~II and Type~IV) rapidly increases the computation time. However, in simulation studies, $6$ degrees of freedom shows to be sufficient for a commonly modelled smooth temperature-mortality curve. Moreover, the penalization order can be lowered from $2$ to $1$, which also allows for a wigglier curve, without the trade-off of an increased computation time. Secondly, in our model, the spatial structure on the DLNM cross-basis is governed by a single variance and correlation parameter, common to all cross-basis coefficients. One might argue that different hyperparameters for the different coefficients would offer greater flexibility. Nonetheless, in our simulation study, our model consistently demonstrated good performance. Moreover, comparison with the INLA version of \cite{Quijal-Zamorano2025}, allowing for a parameter-specific BYM2 structure, does not show any accuracy gain compared to our methods (Table S3). Lastly, the model relies on an empirical Bayes estimate of the hyperparameters, ignoring uncertainty in their posterior distribution. However, since coverage has been shown to be sufficiently high, this limitation does not seem to substantially influence the results.

In summary, our model addresses several key limitations of existing methods in estimating spatially-varying DLNMs. In contrast to the two-stage meta-analysis approach, we allow borrowing strength from neighbouring regions in a one-stage method, possibly in a spatially structured way. Moreover, the use of penalized splines increases flexibility in estimating the exposure-lag-response relationship, compared to the method of \cite{Quijal-Zamorano2024}. Lastly, our method shows a remarkable computational gain in comparison to earlier proposed methods \citep{Economou2025, Quijal-Zamorano2024, Quijal-Zamorano2025}. These advantages make our proposed extension particularly well-suited for the analysis of small area count data, where the spatial differences in exposure-response relationships are of interest.

\section*{Supplementary materials}
\textbf{Online Appendix}: Contains additional simulation studies, tables and figures. (Supplementary.pdf)\\

% \textbf{Data and R scripts}: Data and R codes used to generate the results are available on GitHub (\href{Github link}{\color{blue}{Github link)}}

\section*{Acknowledgements}
The computational resources and services used in this work were provided by the VSC (Flemish Supercomputer Center), funded by the Research Foundation - Flanders (FWO) and the Flemish Government - department EWI.

\section*{Funding}
T.N. gratefully acknowledges funding from Research Foundation - Flanders (grant no. G0A3M24N).
\section*{Competing Interest Statement}
The authors have declared no competing interest.

\newpage
\newpage
\bibliographystyle{apalike}
\bibliography{References}

\end{document}

\maketitle
\section{Laplace approximation and hyperparameter optimization}
\subsection{Laplace approximation}
The Laplace approximation to the posterior distribution $p(\boldsymbol{\xi}| \boldsymbol{\bar{\lambda}}, \boldsymbol{\bar{\tau}}, \boldsymbol{\bar{\rho}},\phi; \mathcal{D})$ requires the estimation of the posterior mode. To this end, the Newton-Raphson algorithm is used, relying on the gradient and Hessian of $\log(p(\boldsymbol{\xi}| \boldsymbol{\bar{\lambda}}, \boldsymbol{\bar{\tau}}, \boldsymbol{\bar{\rho}},\phi; \mathcal{D}))$. From the theory of generalized linear models, these can easily be obtained analytically. For the Poisson log-likelihood, the gradient and Hessian are given by $\nabla_{\boldsymbol{\xi}}\log  \mathcal{L}(\boldsymbol{\xi},\boldsymbol{\bar{\lambda}}, \boldsymbol{\bar{\tau}}, \boldsymbol{\bar{\rho}},\phi; \mathcal{D})=X^{\top}(\boldsymbol{y}-\boldsymbol{\mu})$ and $\nabla^2_{\boldsymbol{\xi}}\log  \mathcal{L}(\boldsymbol{\xi},\boldsymbol{\bar{\lambda}}, \boldsymbol{\bar{\tau}}, \boldsymbol{\bar{\rho}},\phi; \mathcal{D})=-X^{\top}\mathcal{V}X$ respectively, with $\mathcal{V}$ a diagonal matrix with entries $\mu_{t,j}$. For a negative binomial model, the gradient can be written as $\nabla_{\boldsymbol{\xi}}\log  \mathcal{L}(\boldsymbol{\xi},\boldsymbol{\bar{\lambda}}, \boldsymbol{\bar{\tau}}, \boldsymbol{\bar{\rho}},\phi; \mathcal{D})=X^{\top}\mathcal{W}(\boldsymbol{y}-\boldsymbol{\mu})$ with $\mathcal{W}$ a diagonal matrix with entries $\frac{\phi}{\mu_{t,j} + \phi}$. Moreover, the Hessian can be written as $\nabla^2_{\boldsymbol{\xi}}\log  \mathcal{L}(\boldsymbol{\xi},\boldsymbol{\bar{\lambda}}, \boldsymbol{\bar{\tau}}, \boldsymbol{\bar{\rho}},\phi; \mathcal{D})= X^T\mathcal{M}X$, with $\mathcal{M}$ a diagonal matrix with entries $\frac{-\mu_{t,j}\phi(y_{t,j}+\phi)}{(\mu_{t,j}+\phi)^2}$. Therefore, the gradient and Hessian of the log-posterior can easily be calculated as $\nabla_{\boldsymbol{\xi}}\log p(\boldsymbol{\xi}| \boldsymbol{\bar{\lambda}}, \boldsymbol{\bar{\tau}}, \boldsymbol{\bar{\rho}},\phi; \mathcal{D})=\nabla_{\boldsymbol{\xi}}\log  \mathcal{L}(\boldsymbol{\xi},\boldsymbol{\bar{\lambda}}, \boldsymbol{\bar{\tau}}, \boldsymbol{\bar{\rho}},\phi; \mathcal{D})-Q\boldsymbol{\xi}$ and $\nabla^2_{\boldsymbol{\xi}}\log p(\boldsymbol{\xi}| \boldsymbol{\bar{\lambda}}, \boldsymbol{\bar{\tau}}, \boldsymbol{\bar{\rho}},\phi; \mathcal{D})=\nabla^2_{\boldsymbol{\xi}}\log  \mathcal{L}(\boldsymbol{\xi},\boldsymbol{\bar{\lambda}}, \boldsymbol{\bar{\tau}}, \boldsymbol{\bar{\rho}},\phi; \mathcal{D})-Q$ respectively.

\subsection{Hyperparameter optimization}
The joint posterior distribution of all hyperparameters, including precision parameters $\bar{\boldsymbol{\tau}}$, penalty parameters $\bar{\boldsymbol{\lambda}}$, correlation parameters $\bar{\boldsymbol{\rho}}$, dispersion parameter $\phi$ and all other hyperparameters $\boldsymbol{\delta}_{\boldsymbol{\tau}}$ and $\boldsymbol{\delta}_{\boldsymbol{\lambda}}$, is given by:
$$p(\bar{\boldsymbol{\lambda}},\bar{\boldsymbol{\tau}},\bar{\boldsymbol{\rho}}, \phi,\boldsymbol{\delta}_{\boldsymbol{\lambda}}, \boldsymbol{\delta}_{\boldsymbol{\tau}}|\mathcal{D}) \propto \frac{\mathcal{L}(\boldsymbol{\xi},\boldsymbol{\bar{\lambda}}, \boldsymbol{\bar{\tau}}, \boldsymbol{\bar{\rho}},\phi; \mathcal{D})p(\boldsymbol{\xi}| \boldsymbol{\bar{\lambda}}, \boldsymbol{\bar{\tau}}, \boldsymbol{\bar{\rho}},\phi)p(\boldsymbol{\bar{\lambda}}|\boldsymbol{\delta}_{\boldsymbol{\lambda}})p(\boldsymbol{\delta}_{\boldsymbol{\lambda}})p(\boldsymbol{\bar{\tau}}|\boldsymbol{\delta}_{\boldsymbol{\tau}})p(\boldsymbol{\delta}_{\boldsymbol{\tau}})p(\boldsymbol{\bar{\rho}})p(\phi)}{p(\boldsymbol{\xi}| \boldsymbol{\bar{\lambda}}, \boldsymbol{\bar{\tau}}, \boldsymbol{\bar{\rho}},\phi; \mathcal{D})}.$$
Following \cite{rue2009}, the posterior $p(\boldsymbol{\xi}| \boldsymbol{\bar{\lambda}}, \boldsymbol{\bar{\tau}}, \boldsymbol{\bar{\rho}},\phi; \mathcal{D})$ can be replaced by its approximation $\widetilde{p}_G(\boldsymbol{\xi}| \boldsymbol{\bar{\lambda}}, \boldsymbol{\bar{\tau}}, \boldsymbol{\bar{\rho}},\phi; \mathcal{D})$, and by evaluating $\boldsymbol{\xi}$ at the estimated posterior mode $\hat{\boldsymbol{\xi}}$. Note that the determinant $|Q|^{1/2}$ in $p(\boldsymbol{\xi}| \boldsymbol{\bar{\lambda}}, \boldsymbol{\bar{\tau}}, \boldsymbol{\bar{\rho}},\phi)$ can be written as $|Q|^{1/2} \propto |\Omega_\beta|^{1/2}\times |V_{\theta}|^{1/2}\times|V_{\theta_{\mathcal{J}}}|^{1/2}\times |G|^{1/2}$ and that $\widetilde{p}_G(\boldsymbol{\xi}| \boldsymbol{\bar{\lambda}}, \boldsymbol{\bar{\tau}}, \boldsymbol{\bar{\rho}},\phi; \mathcal{D})\bigr|_{\boldsymbol{\xi}=\widehat{\boldsymbol{\xi}}} \propto |\Sigma|^{-1/2}$. Hence, the posterior of the hyperparameters can be rewritten as:
\begin{align*}
    p(\bar{\boldsymbol{\lambda}},\bar{\boldsymbol{\tau}},\bar{\boldsymbol{\rho}}, \phi,\boldsymbol{\delta}_{\boldsymbol{\lambda}}, \boldsymbol{\delta}_{\boldsymbol{\tau}}|\mathcal{D}) & \propto \mathcal{L}(\boldsymbol{\xi},\boldsymbol{\bar{\lambda}}, \boldsymbol{\bar{\tau}}, \boldsymbol{\bar{\rho}},\phi; \mathcal{D}) \times |V_{\theta}|^{1/2}\times|V_{\theta_{\mathcal{J}}}|^{1/2}\times |G|^{1/2} \times \exp\biggl({-\frac{1}{2}}\hat{\boldsymbol{\xi}}^\top Q \hat{\boldsymbol{\xi}}\biggl)\\
    & \times \prod_{\lambda_i \in \bar{\boldsymbol{\lambda}}}{\lambda_i^{\nu/2-1}\delta_{\lambda_i}^{(\frac{\nu}{2}+a-1)}\exp(-\delta_{\lambda_i}(b+\frac{\nu}{2}\lambda_i))} \\
    &\times \prod_{\tau_i \in \bar{\boldsymbol{\tau}}}{\tau_i^{\nu/2-1}\delta_{\tau_i}^{(\frac{\nu}{2}+a-1)}\exp(-\delta_{\tau_i}(b+\frac{\nu}{2}\tau_i))} \\
    & \times \prod_{\rho_i \in \bar{\boldsymbol{\rho}}}{\rho_i^{-\frac{1}{2}}(1-\rho_i)^{-\frac{1}{2}}}\\
    & \times |\hat{\Sigma}|^{1/2} p(\phi). 
\end{align*}
The hyperparameters $\boldsymbol{\delta_{\lambda}}$ and $\boldsymbol{\delta_{\tau}}$ can be integrated out, to obtain:
\begin{align*}
    p(\bar{\boldsymbol{\lambda}},\bar{\boldsymbol{\tau}},\bar{\boldsymbol{\rho}}, \phi|\mathcal{D})& = \int_0^\infty \cdots \int_0^\infty p(\bar{\boldsymbol{\lambda}},\bar{\boldsymbol{\tau}},\bar{\boldsymbol{\rho}}, \phi,\boldsymbol{\delta}_{\boldsymbol{\lambda}}, \boldsymbol{\delta}_{\boldsymbol{\tau}}|\mathcal{D}) d\boldsymbol{\delta}_{\boldsymbol{\lambda}} \, d\boldsymbol{\delta}_{\boldsymbol{_{\tau}}}\\
    &\propto \mathcal{L}(\boldsymbol{\xi},\boldsymbol{\bar{\lambda}}, \boldsymbol{\bar{\tau}}, \boldsymbol{\bar{\rho}},\phi; \mathcal{D}) \times |V_{\theta}|^{1/2}\times|V_{\theta_{\mathcal{J}}}|^{1/2}\times |G|^{1/2} \times \exp\biggl({-\frac{1}{2}}\hat{\boldsymbol{\xi}}^\top Q \hat{\boldsymbol{\xi}}\biggl) \\
    & \times \prod_{\lambda_i \in \bar{\boldsymbol{\lambda}}}{\lambda_i^{\nu/2-1}(b+\frac{\nu}{2}\lambda_i)^{-(\frac{\nu}{2}+a)}} \times \prod_{\tau_i \in \bar{\boldsymbol{\tau}}}{\tau_i^{\nu/2-1}(b+\frac{\nu}{2}\tau_i)^{-(\frac{\nu}{2}+a)}} \\
    &\times |\hat{\Sigma}|^{1/2} p(\phi)\times \prod_{\rho_i \in \bar{\boldsymbol{\rho}}}{\rho_i^{-\frac{1}{2}}(1-\rho_i)^{-\frac{1}{2}}}.\\
\end{align*}
For numerical stability, we use a log-transformation of the penalty vector $\boldsymbol{\psi_{\bar{\lambda}}} = \log(\bar{\boldsymbol{\lambda}})$. Similarly, for the precision parameters, we use $\boldsymbol{\psi_{\bar{\tau}}} = \log(\bar{\boldsymbol{\tau}})$ and for the dispersion parameter, in case of the negative binomial model, we use $\psi_\phi = \log(\phi)$. For the correlation parameters $\bar{\boldsymbol{\rho}}$ of the Leroux model, if included, we use the transformation $\boldsymbol{\psi_{\bar{\rho}}} = \log\bigl(\frac{\bar{\boldsymbol{\rho}}}{1-\bar{\boldsymbol{\rho}}}\bigl)$. Taking into account the Jacobians of the transformations, the log-posterior becomes: %Note that if $p(\phi) \propto \frac{1}{\phi}$ and taking into account that the Jacobian of the transformation $\psi_\phi$ is equal to $\exp(\psi_\phi)$, we obtain $p(\psi_\phi) = \frac{1}{\exp(\psi_\phi)}\exp(\psi_\phi)=1$. 
\begin{align*}
    \log p(\boldsymbol{\psi_{\bar{\lambda}}},\boldsymbol{\psi_{\bar{\tau}}},\boldsymbol{\psi_{\bar{\rho}}}, \psi_\phi|\mathcal{D}) &\propto \log \mathcal{L}(\boldsymbol{\xi},\boldsymbol{\psi_{\bar{\lambda}}},\boldsymbol{\psi_{\bar{\tau}}},\boldsymbol{\psi_{\bar{\rho}}}, \psi_\phi; \mathcal{D}) + \frac{1}{2}\log|V_\theta| + \frac{1}{2}\log|V_{\theta_{\mathcal{J}}}| + \frac{1}{2}\log|G|  \\
    & - \frac{1}{2}{\widehat{\boldsymbol{\xi}}}^{\top} Q \widehat{\boldsymbol{\xi}}+{\frac{1}{2}} \log |\widehat{\Sigma}|+\sum_{\psi_{\rho_i} \in \boldsymbol{\psi_{\bar{\rho}}}}{\biggl(\frac{1}{2}\psi_{\rho_i}-\log(1+\exp(\psi_{\rho_i})\biggl)} \\
    & +\sum_{\psi_{\lambda_i} \in \boldsymbol{\psi_{\bar{\lambda}}}}{\biggl(\frac{\nu}{2}\psi_{\lambda_i}- \left(\frac{\nu}{2}+a\right) \left(\log(b+\frac{\nu}{2}\exp(\psi_{\lambda_i}))\right)\biggl)} \\
    & +\sum_{\psi_{\tau_i} \in \boldsymbol{\psi_{\bar{\tau}}}}{\biggl(\frac{\nu}{2}\psi_{\tau_i}- \left(\frac{\nu}{2}+a\right) \left(\log(b+\frac{\nu}{2}\exp(\psi_{\tau_i}))\right)\biggl)}.
\end{align*}
By maximizing this analytical expression for $\log p(\boldsymbol{\psi_{\bar{\lambda}}},\boldsymbol{\psi_{\bar{\tau}}},\boldsymbol{\psi_{\bar{\rho}}}, \psi_\phi|\mathcal{D})$, we obtain the maximum a posterior estimate of our hyperparameters $\boldsymbol{\psi_{\bar{\lambda}}},\boldsymbol{\psi_{\bar{\tau}}},\boldsymbol{\psi_{\bar{\rho}}}$ and $\psi_\phi$. This estimate can then be used to approximate the marginal posterior of $\boldsymbol{\xi}$, ensuring fast inference compared to MCMC based approaches.

\section{Simulation study}
\subsection{Simulation set-up}
Consider the smooth function:
\begin{align*}
    f_j\cdot w_j(x_{t-l,j},l) = 0.1 \times \sum_{p=1}^{5}\delta_{pj}(x_{t-l,j}-5)^{p-1}w_j(x_{t-l},l)
\end{align*}
with
\begin{align*}
    w_j(x,l) = \begin{cases}
\exp\!\left(-\dfrac{l}{d_j}\right), & x \ge 5, \\[6pt]
4\,\Phi_{2,2}(l), & x < 5.
\end{cases}
\end{align*}
with $\Phi_{m,s}$ a normal density function with mean $m$ and standard deviation $s$. Denote by $\delta = (0.211881,0.1406585,-0.0982663,0.0153671,-0.0006265)^\top$ the baseline coefficients. Then $\delta_{pj} = 1.5\delta_p(1+\Delta_{pj})$ where $\Delta_{pj}$ is simulated, for all $j=1,\ldots,73$, from a Leroux distribution with parameter-specific variance $\sigma_{p}^2 $ and correlation parameter $\rho$. The values of the variances $\sigma_p^2 $ and correlation $\rho$ depend on the assumed setting. In the first setting, the variances are given by $\boldsymbol{\sigma^2} = (0.0001,0.0005,0.00025,0.000005,0.000001)^\top$ and $\rho = 0.95$. In the second setting, the variances increase to $\boldsymbol{\sigma^2} = (0.001,0.005,0.0025,0.00005,0.00001)^\top$, but the correlation is still equal to $\rho = 0.95$. In the third setting, the variance vector is equal to the second setting but the correlation parameter is now equal to $\rho = 0.05$. The denominator $d_j$ is equal to $d_j = 4 + r_j$, with $r_j$ simulated from a Leroux distribution with variance equal to $1$ and the setting-specific $\rho$ parameter.

\subsection{Simulation metrics}
In each of the $250$ simulations and for each method, we summarized the RMSE as:
\begin{align*}
    \text{RMSE} = \sqrt{\frac{1}{ 73 \times 41 \times 9}\sum_{j=1}^{73}\sum_{x=0}^{10}\sum_{l=0}^{8}{\biggl(\frac{1}{250}\sum_{s = 1}^{250}\bigl(\log\bigl(\widehat{RR}^{s}_{x,x_0,j}(l)\bigl) - \log\bigl(RR^{\text{true}}_{x,x_0,j}(l)\bigl)\bigl)^2\biggl) }},
\end{align*}
for the lag-specific RR and 
\begin{align*}
    \text{RMSE} = \sqrt{\frac{1}{ 73 \times 41 \times 9}\sum_{j=1}^{73}\sum_{x=0}^{10}{\biggl(\frac{1}{250}\sum_{s = 1}^{250}\bigl(\log\bigl(\widehat{RR}^{\text{overall},s}_{x,x_0,j}\bigl) - \log\bigl(RR^{\text{overall,true}}_{x,x_0,j}\bigl)\bigl)^2\biggl) }},
\end{align*}
for the overall cumulative RR. The coverage of the $95\%$ credible intervals was caclulated as:
\begin{align*}
    \text{coverage} = \frac{1}{ 73 \times 41 \times 9}\sum_{j=1}^{73}\sum_{x=0}^{10}\sum_{l=0}^{8}{\biggl(\frac{1}{250}\sum_{s = 1}^{250}\bigl( \log\bigl(RR^{\text{true}}_{x,x_0,j}(l)\bigl) \in \text{CI}_s\bigl)\biggl) },
\end{align*}
and
\begin{align*}
    \text{coverage} = \frac{1}{ 73 \times 41 \times 9}\sum_{j=1}^{73}\sum_{x=0}^{10}\sum_{l=0}^{8}{\biggl(\frac{1}{250}\sum_{s = 1}^{250}I\bigl(\log\bigl(RR^{\text{true}}_{x,x_0,j}(l)\bigl ) \in \text{CI}_s\bigl)\biggl) },
\end{align*}
where $\text{CI}_s$ is the relevant credible interval, calculated using the posterior distribution obtained from simulation $s$.

\subsection{Additional simulation study}

\subsubsection{Negative binomial}
In this simulation study, we used the same simulation settings as in the main paper, but in a negative binomial setting. We simulated from a negative binomial model with dispersion parameter equal to $5$ ($\text{Var}(y_{t,j}) = \mu_{t,j}+\frac{\mu_{t,j}^2}{5}$). The results can be found in Table \ref{tab:simulation_study_small_NB} and Table \ref{tab:simulation_study_large_NB} for the small and large area setting respectively. Similar observations can be made as in the Poisson case, although we see that the models are more difficult to estimate due to the additional variability of the negative binomial model. However, penalization seems even more beneficial in this case of overdisperion.

\begin{table}[h]
\caption{Simulation results for small area setting with a negative binomial model: RMSE of lag-specific RR (RMSE RR), RMSE of cumulative RR, over lag $0-8$, (RMSE RR overall), coverage of lag-specific RR (cov RR),  coverage of cumulative RR (cov RR overall), AUC of top $10\%$ and AUC of top $25\%$ high risk areas. The computation time is reported in seconds.}
\label{tab:simulation_study_small_NB}
\centering
\begin{adjustbox}{width=\linewidth} % Optional for scaling if necessary
\begin{tabular}{llccccccc}
\hline
 \textbf{Setting} &  \textbf{Method} & \textbf{time} & \textbf{cov RR} &\textbf{cov RR} & \textbf{RMSE RR} &\textbf{RMSE RR} & \textbf{AUC $10\%$} & \textbf{AUC $25\%$} \\ 
  &   &  &  & \textbf{overall} &  &\textbf{overall} & & \\ \hline
   \multirow{6}{*}{Setting 1} & Type I  & $43.06$ & $0.9968$ & $0.9933$ & $0.0223$& $0.1005$ & $0.5950$ & $0.5687$\\
                          & Type II  & $72.76$ & $0.9838$ & $0.9870$ & $0.0138$& $0.0673$ & $0.6122$ & $0.5816$ \\
                          & Type III   & $44.12$ & $0.9847$ & $0.9870$& $0.0284$ & $0.1295$ & $0.6289$ & $0.5916$ \\
                          & Type IV  & $130.83$ & $0.9924$ & $0.9942$& $0.0133$ & $0.0655$ & $0.6566$ & $0.6132$ \\
                          & Common  & $43.07$ & $0.9672$ & $0.9119$ & $0.0137$& $0.0646$ & $0.5000$ & $0.5000$ \\
                          & Meta  & $27.20$ & $0.9861$ & $0.9796$ & $0.0353$&  $0.1280$& $0.5886$ & $0.5579$\\

\hline
 \multirow{6}{*}{Setting 2} & Type I   & $43.68$ & $0.9968$& $0.9922$ & $0.0235$ & $0.1140$ & $0.6555$ & $0.6350$  \\
                          & Type II  & $73.50$& $0.9596$ & $0.9220$ & $0.0156$& $0.0860$ & $0.7202$ & $0.7016$  \\
                          & Type III   & $44.65$ & $0.9804$& $0.9657$ & $0.0300$& $0.1454$ & $0.7230$ & $0.6800$   \\
                          & Type IV   & $131.53$ & $0.9850$& $0.9799$ & $0.0145$ & $0.0776$ & $0.8131$ & $0.7553$   \\
                          & Common   & $43.26$& $0.9214$ & $0.6912$ & $0.0163$& $0.0896$& $0.5000$  & $0.5000$  \\
                          & Meta   & $27.74$ & $0.9816$& $0.9615$ & $0.0366$& $0.1418$ & $0.6329$ & $0.6094$  \\

\hline
 \multirow{6}{*}{Setting 3} & Type I   & $43.99$ & $0.9961$ &$0.9898$ & $0.0255$& $0.1329$ & $0.7324$ & $0.6849$\\
                          & Type II   & $74.16$ & $0.9321$ &$0.8490$ &$0.0181$ & $0.1139$ & $0.8232$ & $0.7665$ \\
                          & Type III   & $45.31$ & $0.9692$ & $0.9197$& $0.0309$& $0.1600$ & $0.7440$ & $0.6907$ \\
                          & Type IV  & $134.14$ & $0.9664$ &$0.9317$ & $0.0174$& $0.1076$ & $0.8119$ & $0.7405$\\
                          & Common   & $43.89$ & $0.8693$ &$0.5747$ & $0.0193$& $0.1216$ & $0.5000$ & $0.5000$\\
                          & Meta   & $28.03$ & $0.9779$ &$0.9546$ & $0.0377$& $0.1526$ & $0.7322$ &$0.6857$ \\

\hline

\end{tabular}
\end{adjustbox}
\end{table}

\begin{table}[h]
\caption{Simulation results for large area setting with a negative binomial model: RMSE of lag-specific RR (RMSE RR), RMSE of cumulative RR, over lag $0-8$, (RMSE RR overall), coverage of lag-specific RR (cov RR),  coverage of cumulative RR (cov RR overall), AUC of top $10\%$ and AUC of top $25\%$ high risk areas. The computation time is reported in seconds.}
\label{tab:simulation_study_large_NB}
\centering
\begin{adjustbox}{width=\linewidth} % Optional for scaling if necessary
\begin{tabular}{llccccccc}
\hline
 \textbf{Setting} &  \textbf{Method} & \textbf{time} & \textbf{cov RR} &\textbf{cov RR} & \textbf{RMSE RR} &\textbf{RMSE RR} & \textbf{AUC $10\%$} & \textbf{AUC $25\%$} \\ 
  &   &  &  & \textbf{overall} &  &\textbf{overall} & & \\ \hline
   \multirow{6}{*}{Setting 1} & Type I & $32.01$ & $0.9892$ & $0.9777$&$0.0097$ & $0.0493$ & $0.6735$ & $0.6442$\\
                          & Type II  & $58.41$ & $0.9883$ & $0.9869$& $0.0079$ & $0.0446$ & $0.7107$ & $0.6796$\\
                          & Type III   & $33.74$ & $0.9830$ & $0.9671$& $0.0101$ & $0.0510$ & $0.7055$ & $0.6696$ \\
                          & Type IV  & $116.65$ & $0.9920$ & $0.9893$& $0.0081$ & $0.0458$ & $0.7294$ &  $0.6915$\\
                          & Common  & $32.44$ & $0.9369$ &$0.7560$ & $0.0081$ & $0.0444$ & $0.5000$  & $0.5000$ \\
                          & Meta  & $28.04$ & $0.9841$ & $0.9628$& $0.0180$  & $0.0648$ & $0.6562$ & $0.5994$ \\

\hline
 \multirow{6}{*}{Setting 2} & Type I   & $32.10$ &  $0.9821$ & $0.9588$ & $0.0120$ & $0.0677$ & $0.8302$ & $0.8143$ \\
                          & Type II   & $58.73$ & $0.9707$ & $0.9534$ & $0.0095$& $0.0576$ & $0.9052$  & $0.8907$ \\
                          & Type III   & $33.88$ & $0.9738$ & $0.9395$ & $0.0124$ & $0.0686$ & $0.8518$ & $0.8261$ \\
                          & Type IV   & $117.59$ & $0.9880$ & $0.9811$& $0.0095$ & $0.0555$ & $0.8992$ & $0.8784$ \\
                          & Common   & $32.44$ & $0.7921$ & $0.4187$& $0.0116$& $0.0749$  & $0.5000$  & $0.5000$ \\
                          & Meta   & $27.37$ & $0.9742$ & $0.9536$& $0.0187$ & $0.0717$ & $0.8437$ & $0.8294$ \\

\hline
 \multirow{6}{*}{Setting 3} & Type I   & $33.27$ & $0.9739$ & $0.9371$ & $0.0144$& $0.0863$ & $0.8855$ & $0.8553$\\
                          & Type II   & $61.68$ & $0.9519$ & $0.9141$&$0.0133$ & $0.0747$ & $0.9635$ & $0.9280$ \\
                          & Type III   & $35.76$ & $0.9485$ & $0.8799$& $0.0149$& $0.0900$ & $0.8877$ & $0.8564$ \\
                          & Type IV  & $121.98$ & $0.9771$ &$0.9600$ & $0.0113$ & $0.0716$ & $0.9463$ & $0.9100$\\
                          & Common   & $33.73$ & $0.7139$ & $0.3657$& $0.0150$ & $0.1060$ & $0.5000$ & $0.5000$\\
                          & Meta   & $28.67$ & $0.9719$ & $0.9577$& $0.0191$& $0.0742$ & $0.9338$ & $0.8972$\\

\hline

\end{tabular}
\end{adjustbox}
\end{table}

\subsubsection{Comparison to existing methods}
In this simulation study, we briefly compared the results obtained from our proposed methods, with the results that can be obtained using the method of \cite{Economou2025}, defined using the \texttt{bam} function in the \texttt{mgcv} package, and the results obtained from the method of \cite{Quijal-Zamorano2025}, using \texttt{INLA}. Before discussing the results, we provide a brief overview of both methods.

\subsubsection*{Method of Economou et al.}
In the paper of \cite{Economou2025}, different methods are proposed to model a spatially varying exposure-lag-response curve within a DLNM framework. All models build on tin-plate regression splines (TPRS) and are therefore penalized spline methods. Their first method assumes smooth spatial variability, specifying an interaction between the DLNM cross-basis and the coordinates of the centroids of the area, defined by the longitude (lon) and latitude (lat). This results in the following model:
\begin{align*}
    \log(\mu_{t,j}) = \beta_0+\sum_{h=1}^{H}{\beta_h a_{t,j,h}}+ f(\text{lon}_j, \text{lat}_j) + \sum_{l=0}^{L}{s_1(l,x_{t-l})} + \sum_{l=0}^{L}{s_2(l,x_{t-l}, \text{lon}_j, \text{lat}_j)},
\end{align*}
where $s$ represents a smooth function, constructed as a tensor product of marginal thin-plate regression splines. This way, the function $s_1(l,x_{t-l})$ corresponds to:
\begin{align*}
    \sum_{i=1}^{v_x} \sum_{k=1}^{v_l}{\left(\widetilde{b}_i(x_{t-l,j})\shortsmile{b}_k(l)\right)\theta_{ik}}
\end{align*}
defined earlier. The function $s_2(l,x_{t-l}, \text{lon}_j, \text{lat}_j)$ similarly corresponds to the location-specific deviation, varying smoothly with the coordinates.

Alternatively, they define a spatially-varying DLNM model by specifying an interaction with a Markov Random Field (MRF). The model can then be formulated as:
\begin{align*}
    \log(\mu_{t,j}) = \beta_0+\sum_{h=1}^{H}{\beta_h a_{t,j,h}}+ f(j) + \sum_{l=0}^{L}{s_1(l,x_{t-l})} + \sum_{l=0}^{L}{s_2(l,x_{t-l}, j)},
\end{align*}
where $f(j)$ corresponds to a Markov Random Field spline and $s_2$ is a tensor product between a marginal TPRS in both the exposure and lag dimension and another MRF spline.

Lastly, they propose a model assuming region-specific deviations of a global trend that are unstructured in space. Hence, the model can similarly defined by:
\begin{align*}
    \log(\mu_{t,j}) = \beta_0+\sum_{h=1}^{H}{\beta_h a_{t,j,h}} + \sum_{l=0}^{L}{s_1(l,x_{t-l})} + \sum_{l=0}^{L}{s_2(l,x_{t-l}, j)}+u_j,
\end{align*}
but with $u_j \sim \mathcal{N}(0,\sigma^2)$ an independent random effect and $s_2$ a tensor product between a marginal TPRS in the exposure and lag dimension and a random effect spline.

\subsubsection*{Method of Quijal-Zamorano et al.}
Because of the computational burden of the MCMC approach proposed by \cite{Quijal-Zamorano2024}, we only included a comparison with the INLA version of this method, explained by \cite{Quijal-Zamorano2025}. The model can be written as:
\begin{align*}
    \log(\mu_{t,j}) = \beta_0+\sum_{h=1}^{H}{\beta_h a_{t,j,h}}+ \sum_{i=1}^{v_x} \sum_{k=1}^{v_l}\left(\sum_{l=0}^L\widetilde{b}_i(x_{t-l,j})\shortsmile{b}_k(l) \right)(\theta_{ik}+\theta_{j,ik}),
\end{align*}
using unpenalized splines for both the exposure and lag dimension. Although the original method of \cite{Quijal-Zamorano2024} is defined in terms of a Leroux random effect, this is not straightforward in INLA. Therefore, the INLA implementation makes use of a BYM2 effect \citep{Simpson2017}. In contrast to our model specification, each parameter $\theta_{j,k}$ follows a BYM2 model with its own precision parameter $\tau_k$ and mixing parameter $\phi_k$. Although a spatially structured random effect $u_j$ is originally not included in this model, it can easily be added in INLA, specifying again a BYM2 structure.

\subsubsection*{Simulation results}
These results are based on only $10$ simulations runs, because of the high computational demands. We used Setting~2, defined in the main paper, in both a small and large area setting, simulating from a Poisson model. The reported computation time is the time (in seconds) obtained from a device with an AMD Ryzen(TM) 5 PRO 7530U processor (base frequency 2.00GHz), having six cores (12 threads) and 16GB of RAM. Because of the computational burden of the MRF method of \cite{Economou2025}, we only included their first spatial method (based on longitude and latitude) and their spatially unstructured method in the comparison. The results can be found in Table \ref{tab:simulation_study_comparison}. It can be seen that the INLA method of \cite{Quijal-Zamorano2025} performs very similar to our Type~III, i.e. spatially structured but unpenalized, method. Although, in contrast to our method, \cite{Quijal-Zamorano2025} allows for a different variance and mixing parameter for every cross-basis coefficient, their method does not outperform ours in terms of RMSE, coverage or AUC. Moreover, their method does not allow for the specification of a penalty term and is computationally less efficient, taking about $15$ times more time than our Type~III method. Comparing our method to the unstructured \texttt{bam} specification of \cite{Economou2025}, we can see that their method performs very similar to our Type~II, i.e. penalized but unstructured, method. However, our Type~II method is much more computational efficient, taking about $30-80$ times less time. Although their proposition based on coordinates decreases the computational burden, it also decreases coverage and AUC compared to our spatially structured (Type~III and Type~IV) methods.

\begin{table}[h]
\caption{Simulation results for Setting~2 (small and large area): RMSE of lag-specific RR (RMSE RR), RMSE of cumulative RR, over lag $0-8$, (RMSE RR overall), coverage of lag-specific RR (cov RR),  coverage of cumulative RR (cov RR overall), AUC of top $10\%$ and AUC of top $25\%$ high risk areas. The computation time is reported in seconds.}
\label{tab:simulation_study_comparison}
\centering
\begin{adjustbox}{width=\linewidth} % Optional for scaling if necessary
\begin{tabular}{llccccccc}
\hline
 \textbf{Area} &  \textbf{Method} & \textbf{time} & \textbf{cov RR} &\textbf{cov RR} & \textbf{RMSE RR} &\textbf{RMSE RR} & \textbf{AUC $10\%$} & \textbf{AUC $25\%$} \\ 
  &   &  &  & \textbf{overall} &  &\textbf{overall}  \\ \hline
 \multirow{9}{*}{Small} & Type I  & $38.21$&$0.9947$ &$0.9912$& $0.0159$ & $0.0815$& $0.7038$& $0.6757$ \\
                          & Type II   & $73.71$ & $0.9977$& $0.9921$& $0.0125$ & $0.0755$&  $0.7357$& $0.7001$ \\
                          & Type III  & $47.54$ & $0.9874$& $0.9798$& $0.0152$ &$0.0751$& $0.7844$ & $0.7301$  \\
                          & Type IV   & $149.326$ & $0.9959$& $0.9862$& $0.0114$ & $0.0658$& $0.8149$& $0.7580$ \\
                          & Common   & $42.94$ &$0.8391$ & $0.5011$& $0.0124$ & $0.0781$&$0.5000$ & $0.5000$ \\
                          & Meta  & $32.91$ &$0.9794$& $0.9389$& $0.0297$&$0.1294$&$0.6813$ &$0.6480$  \\
                          & Bam (lat,lon)  & $171.56$&$0.7971$& $0.6020$& $0.0135$ & $0.0791$ & $0.7006$&$0.6525$ \\
                          & Bam (unstructured)   & $5989.70$ &$0.9302$ & $0.9391$ & $0.0123$ &$0.0725$ &$0.7208$& $0.7011$   \\
                          & INLA   & $772.96$ & $0.9836$ & $0.9611$& $0.0156$& $0.0779$&$0.7466$ & $0.6966$ \\

\hline
 \multirow{9}{*}{Large} & Type I  & $32.49$ & $0.9204$&  $0.8497$& $0.0078$& $0.0442$ & $0.9509$ & $0.9422$  \\
                          & Type II   & $69.22$ & $0.9751$ &$0.9527$ & $0.0051$& $0.0264$ & $0.9680$ & $0.9650$  \\
                          & Type III  & $39.37$ & $0.9092$ &$0.8229$ & $0.0070$ & $0.0426$ & $0.9604$ & $0.9491$\\
                          & Type IV   & $143.60$& $0.9718$& $0.9426$& $0.0043$ & $0.0246$ & $0.9745$ & $0.9715$ \\
                          & Common  & $37.33$ & $0.3364$& $0.0830$& $0.0095$& $0.0694$& $0.5000$& $0.5000$ \\
                          & Meta  & $35.41$& $0.8553$& $0.8517$ & $0.0055$ & $0.0341$ & $0.9894$ & $0.9895$ \\
                          & Bam (lat,lon)  & $141.61$ & $0.3953$& $0.0876$ & $0.0094$ & $0.0687$& $0.8107$ & $0.7426$\\
                          & Bam (unstructured)   &$2521.68$ & $0.9001$& $0.8259$& $0.0057$ & $0.0360$ & $0.9680$& $0.9667$ \\
                          & INLA  & $723.33$& $0.8283$& $0.7945$ & $0.0066$ & $0.0413$ & $0.9497$ & $0.9432$\\

\hline

\end{tabular}
\end{adjustbox}
\end{table}

\newpage
\section{Additional figures and tables}

\begin{figure}[H]
    \centering
    \includegraphics[width=1\linewidth]{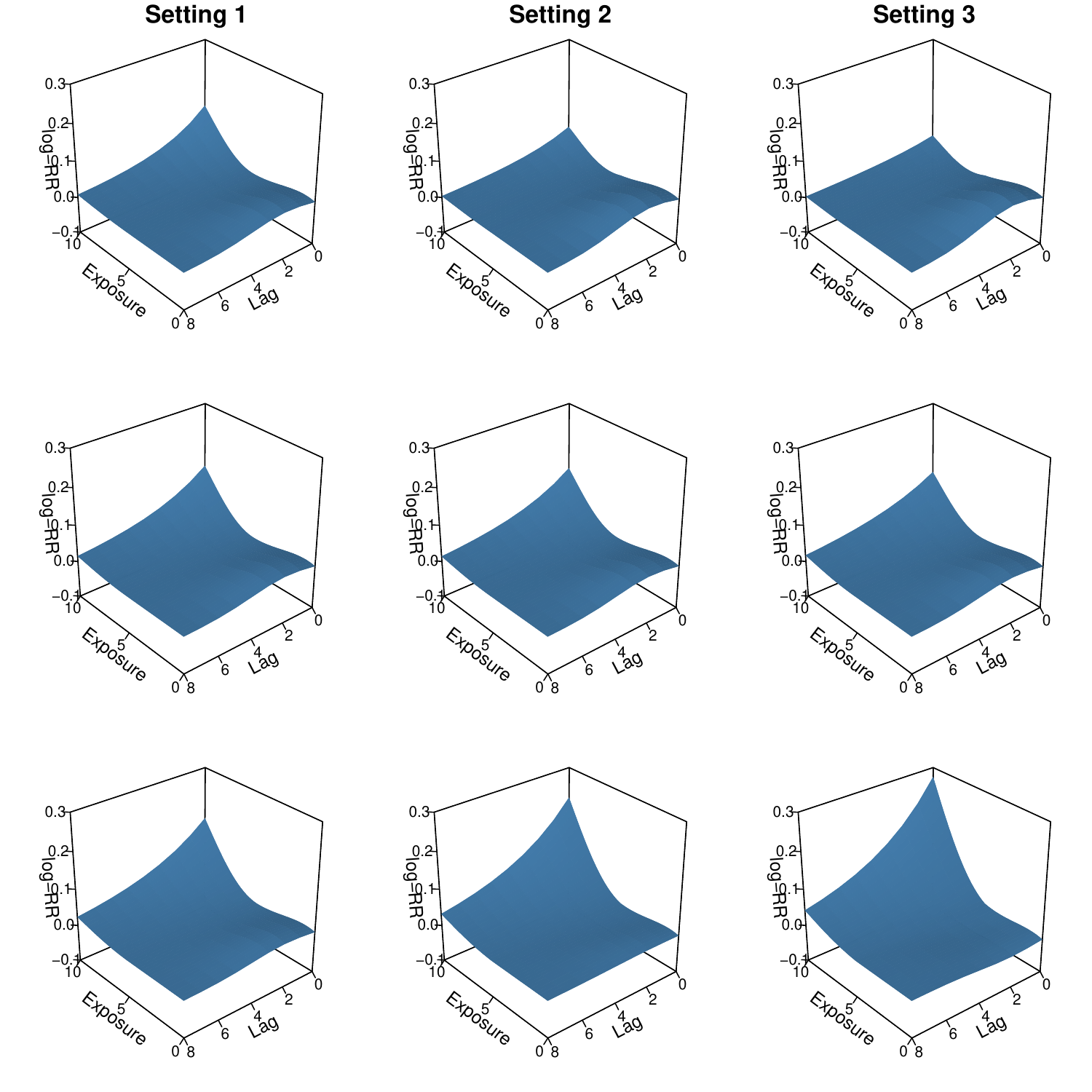}
    \caption{Exposure-lag-response curve for three different areas (rows) in each of the three settings (columns).}
    \label{fig:simulation_lag}
\end{figure}

\begin{figure}[H]
    \centering
    \begin{subfigure}{\textwidth}
        \centering
        \includegraphics[width=\linewidth]{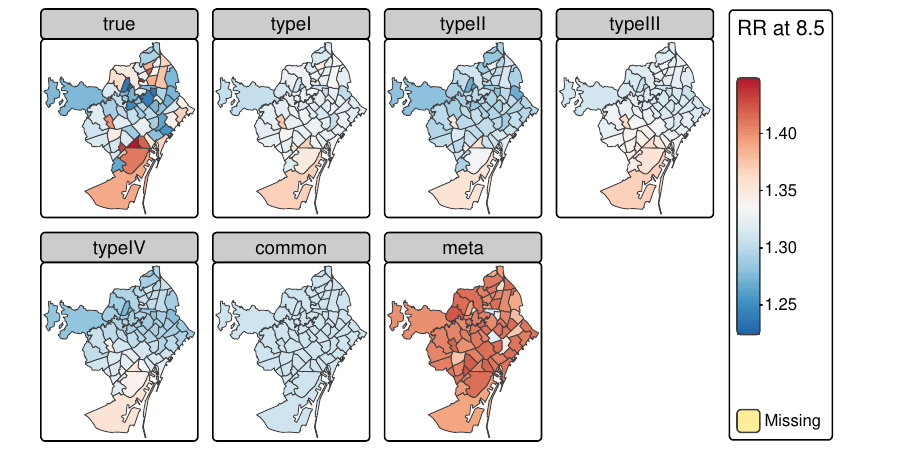}
        \label{fig:setting1_map_small}
        \caption{}
    \end{subfigure}
    \hfill
    \begin{subfigure}{\textwidth}
        \centering
        \includegraphics[width=\linewidth]{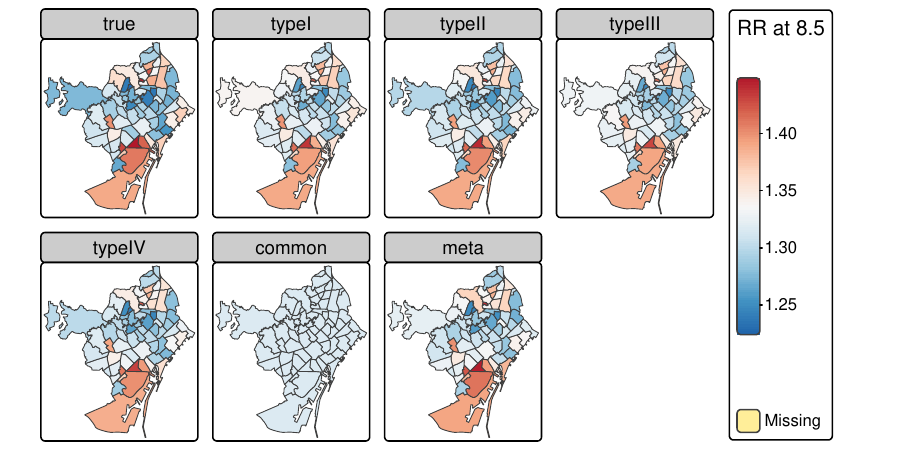}
        \label{fig:setting1_map_large}
         \caption{}
    \end{subfigure}
    \caption{Estimated overall cumulative RR for Setting~1 (averaged over the $250$ simulation runs) in the small area scenario (a) and large area scenario (b).}
    \label{fig:setting1_map}
\end{figure}

\begin{figure}[H]
    \centering
    \begin{subfigure}{\textwidth}
        \centering
        \includegraphics[width=\linewidth]{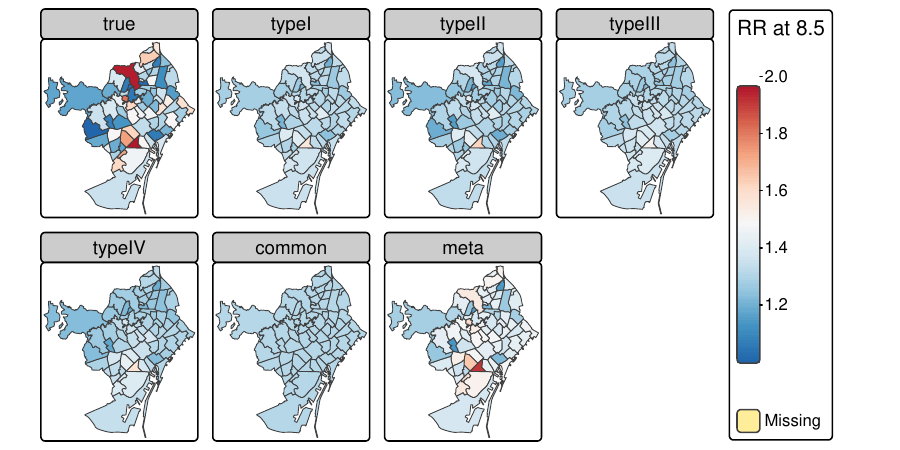}
        \label{fig:setting3_map_small}
        \caption{}
    \end{subfigure}
    \hfill
    \begin{subfigure}{\textwidth}
        \centering
        \includegraphics[width=\linewidth]{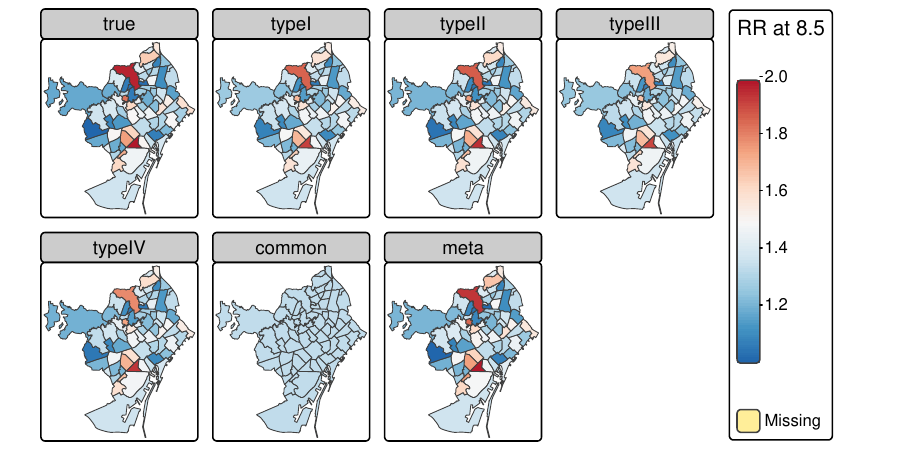}
        \label{fig:setting3_map_large}
         \caption{}
    \end{subfigure}
    \caption{Estimated overall cumulative RR for Setting~3 (averaged over the $250$ simulation runs) in the small area scenario (a) and large area scenario (b).}
    \label{fig:setting3_map}
\end{figure}

\begin{figure}[H]
    \centering
    \includegraphics[width=1\linewidth]{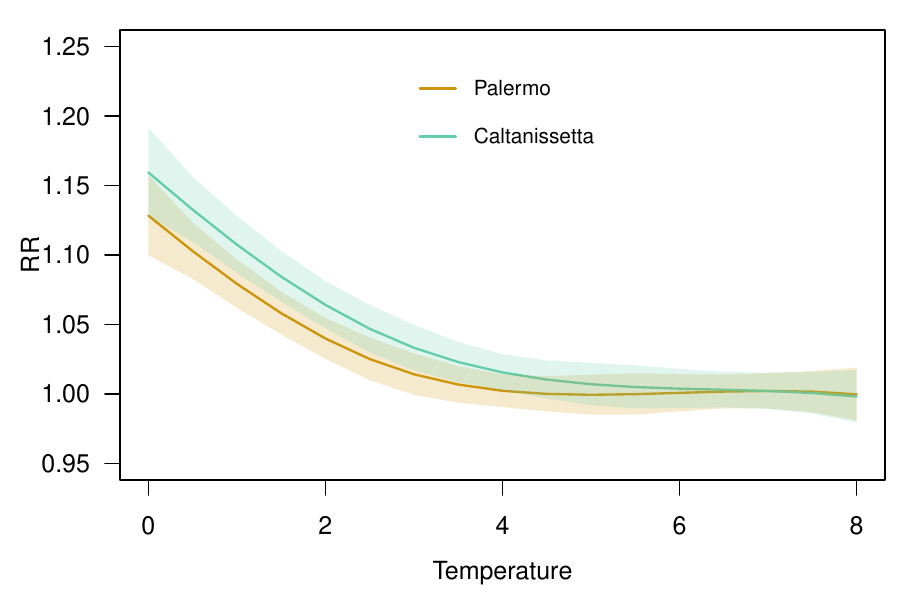}
    \caption{Lag-specific RR at $28$ degrees in Palermo and Caltanissetta, compared to a reference value of $20$ degrees.}
    \label{fig:lag_specific_Sicily}
\end{figure}

\begin{figure}[H]
    \centering
    \includegraphics[width=1\linewidth]{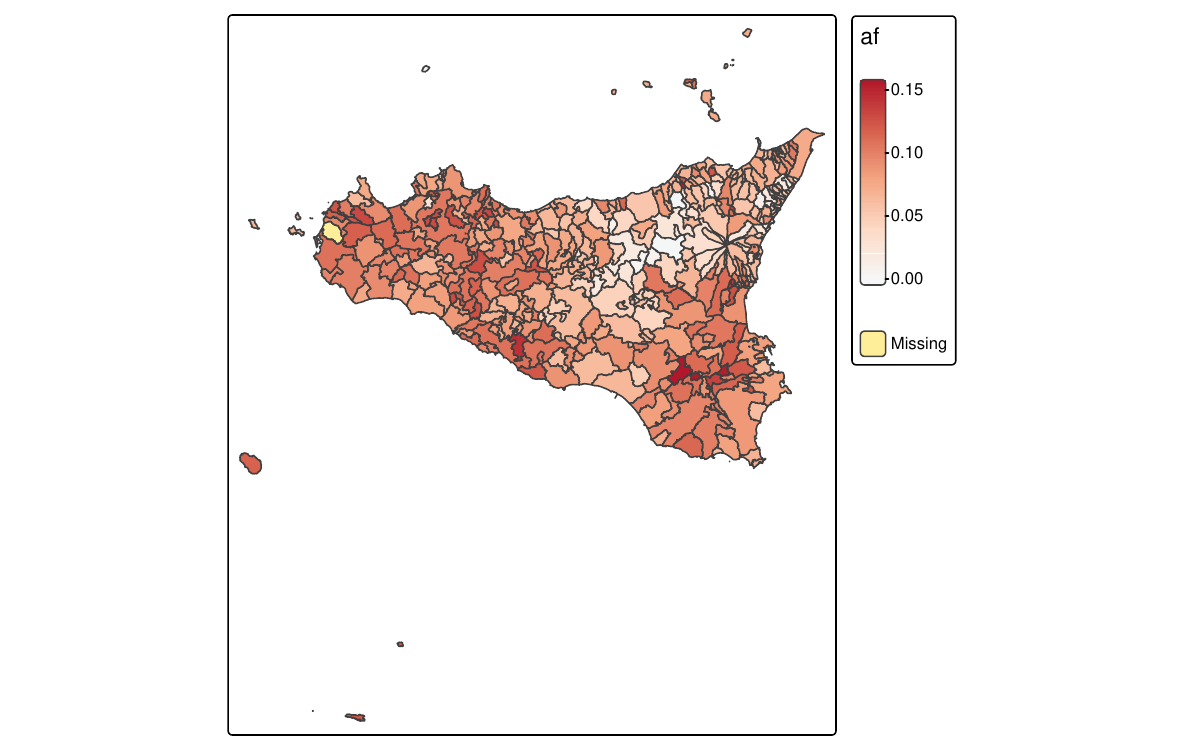}
    \caption{Estimated attributable fraction in $2021$, in every region of Sicily.}
    \label{fig:af_sicily}
\end{figure}

\begin{figure}[H]
    \centering
    \includegraphics[width=1\linewidth]{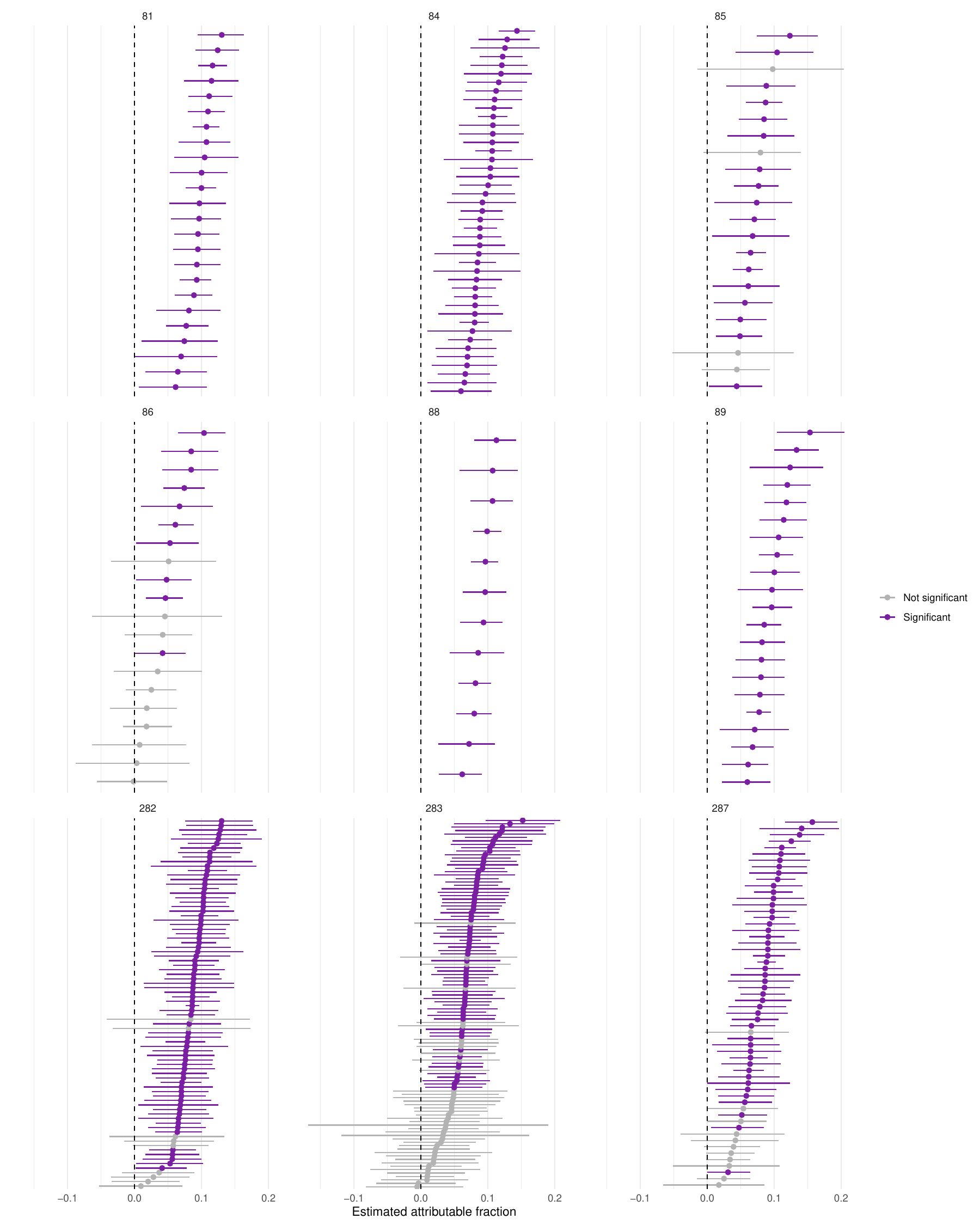}
    \caption{Estimated attributable fraction in $2021$, together with the $95\%$ credible interval, grouped by region.}
    \label{fig:af_caterpillar}
\end{figure}

\begin{table}
    \centering
        \caption{Computation times (in minutes) for the different models, for the Poisson and negative binomial model. The reported time is the time obtained from a device with an AMD Ryzen(TM) 5 PRO 7530U processor (base frequency 2.00GHz), having six cores (12 threads) and 16GB of RAM.}
    \begin{adjustbox}{width=\textwidth}
    \begin{tabular}{r|rrrrr|rrrrr}
        \hline 
        &  \multicolumn{5}{|c|}{Poisson}& \multicolumn{5}{c}{negative binomial} \\
        \hline
        & \multicolumn{1}{|c}{Type I} & Type II & Type III & Type IV & \multicolumn{1}{c|}{Common} & Type I & Type II & Type III & Type IV & Common \\
        \hline
        time & $30.05$ & $44.00$ & $29.45$ & $48.53$& $30.98$ & $41.96$ & $56.31$& $44.28$ & $64.86$ &$43.19$  \\
        \hline
    \end{tabular}
    \end{adjustbox}
    \label{tab:DIC_sicily}
\end{table}

\newpage
\bibliographystyle{apalike}
\bibliography{References}